\documentclass[fleqn,usenatbib]{mnras}

\usepackage{newtxtext,newtxmath}

\usepackage[T1]{fontenc}

\DeclareRobustCommand{\VAN}[3]{#2}
\let\VANthebibliography\thebibliography
\def\thebibliography{\DeclareRobustCommand{\VAN}[3]{##3}\VANthebibliography}

\usepackage{graphicx,subcaption}	
\usepackage{amsmath}	
\newcommand\given[1][]{\:#1\vert\:}
\newcommand{\reviewMoller}[1]{\textcolor{black}{#1}}

\title[Light curve characterisation with mTAN]{Fast and Flexible Characterisation of Astronomical Light Curves Using Multi-Time Attention}

\author[Gondhalekar et al.]{
Y. Gondhalekar,$^{1}$\thanks{E-mail: yash.gondhalekar.25@ucl.ac.uk}
A. Möller,$^{2}$ and
P. Sánchez-Sáez$^{3}$
\\
$^{1}$Department of Physics \& Astronomy, University College London, Gower
Street, London WC1E 6BT, UK\\
$^{2}$Centre for Astrophysics and Supercomputing, Swinburne University of Technology, John St, Hawthorn, VIC 3122, Australia\\
$^{3}$European Southern Observatory, Karl-Schwarzschild-Str. 2, 85748 Garching bei München, Germany
}

\date{Accepted XXX. Received YYY; in original form ZZZ}

\pubyear{\the\year{}}

\begin{document}
\label{firstpage}
\pagerange{\pageref{firstpage}--\pageref{lastpage}}
\maketitle

\begin{abstract}
We present an unsupervised, data-driven framework for rapid characterisation of astronomical photometric time series using a Multi-Time Attention Network. The model learns time-aware latent representations directly from irregular, partial light curves without heavy preprocessing. Through application on ZTF alert data retrieved with Fink, a community alert broker for Rubin LSST, we demonstrate that the model: (i) produces accurate interpolations with small bias (0.01 mag) and scatter (0.1 mag) even for sparse light curves, (ii) learns a temporally distributed latent space correlating with physically meaningful properties (duration, peak time, variability, color) while being robust to unimportant properties such as observed magnitude and number of observations, (iii) separates general SN and AGN samples despite data being heavily dominated by AGNs, and (iv) generalises to unseen classes: The long-period variable and TDE show good interpolation and sensible latent space placement; however, the model cannot capture RRLyrae's $\sim$0.4-0.5 day pulsation period, which is far below our model's chosen two-day temporal resolution. Attention map analysis reveals the capability of multi-time attention to capture local structure. The model is extremely lightweight (a few hundred kilobytes) and has fast inference ($\sim$0.01 and $\sim$$3\times10^{-4}$ s per light curve on CPU and GPU, respectively) that is independent of the number of observations, unlike GP regression. Our approach offers flexible and scalable characterisation, with high relevance in the Rubin LSST era. We discuss future possibilities to incorporate observational uncertainties and symmetries for robustness and forecasting applications for real-time follow-up.
\end{abstract}




\section{Introduction}\label{sec:intro}

New wide-field astronomical surveys such as the Vera C. Rubin Observatory's Legacy Survey of Space and Time \citep[LSST][]{LSST_Ivezic_2019} and the Nancy Grace Roman Space Telescope's High-latitude Time Domain Survey \citep{roman_high_latitude_time_domain} will transform our understanding of transient physics. LSST alone will detect ${\gtrsim}10^6$ supernovae (SNe) annually, which is roughly $100\times$ more than from current surveys combined \citep{Bellm_2019_ztf, PanSTARRS1_Chambers_2016, ATLAS_Tonry_2018, Gaia_Collaboration_2016} and orders of magnitude more than current largest type Ia SN samples \citep{Moller_2022,DES_2024,Moller_2024,Rigault_2025}. LSST's immense type Ia SN sample will help improve cosmological constraints \citep[][]{Graziani2020, Vincenzi_2024}, while two orders of magnitude more superluminous SNe and tidal disruption events (TDEs) will challenge our understanding of their energy sources and physical properties \citep{Villar_2018, Scovacricchi_2016, Bricman2020}. Additionally, tens of millions of AGNs, an order of magnitude more than existing samples, will probe variability from minute to year timescales and help better understand their central engines \citep{LSST_Ivezic_2019}.

Photometric classification is essential as follow-up spectroscopy will be feasible for only a tiny fraction (${\lesssim}1\%$) of the detections from large surveys due to limited telescope resources. Several photometric classification methods have been developed, ranging from those using hand-curated features based on prior astrophysical knowledge \citep[e.g.,][]{D-Isanto2016,Mistry_2024}, extracting features through fitting parametric models to light curves \citep[e.g.,][]{Moller2016, Villar_2019,Sanchez_2021,deSoto_2024,fink_early_SNIa} or related template-fitting methods \citep[e.g.,][]{Sako_2011,Sako_2018}, physical modelling \citep[e.g.,][]{Guillochon_2018,LlamasLanza_2026}, using host galaxy photometry for more rapid follow-ups \citep[e.g.,][]{Kisley_2023} or combining it with light curve data \citep[e.g.,][]{Gagliano_2023}, to non-parametric feature extraction methods using principal component analysis \citep[e.g.,][]{Ishida_2013}, Gaussian process (GP) regression \citep[e.g.,][]{Boone_2019} along with wavelet decomposition \citep[e.g.,][]{Lochner_2016,Alves_2022}.

However, these methods face some limitations. Curating hand-engineered features is time-consuming and biased by our physical understanding of transient phenomena, which may not be optimal for complex subtype classification \citep[e.g., Table 5 of][]{Sanchez_2021,SuperNNova_2020} or for identifying anomalous events \citep{Pruzhinskaya_2026}. Although model parameters are generally interpretable, parametric models suffer from computational expense, the possibility of parameter degeneracies \citep{Villar_2019,SuperNNova_2020}, and model misspecification risks for poorly understood transients or require class-dependent fits. GP regression, though popular for modeling astronomical light curves \citep[][]{Aigrain2023}, suffers from high computational cost and sensitivity to the choice of covariance function, with no universal guidance for handling diverse transient types while keeping low computational costs \citep[see also section 5.5 of][]{Boone_2019}\footnote{Faster linear-time variants of GP with physically motivated covariance function have been proposed \citep[e.g.,][]{Foreman-Mackey_2017}; however, these generally assume one-dimensional settings.}.

These limitations motivate the choice of data-driven approaches using deep neural networks that learn `latent' features from light curves \citep[see also][]{Demianenko_2023}. Several such approaches have been proposed \citep[e.g.,][]{Charnock_2017,RAPID_2019,Pelican_2019,SuperNNova_2020,ParSNIP_2021,Sanchez-Saez_CSAGN,Fraga_2024}, including those aiming to forecast future photometric behavior based on a few observations for optimising the follow-up strategy \citep{Sravan_2020} or those using a combination of image data, alert photometry, and host information \citep{Sheng_2024}. Attention-based architectures have shown particular promise \citep[e.g.,][]{Pimentel_2023,Allam_2024}. For example, \citet{Pimentel_2023} demonstrated especially beneficial for classification using a few observations on real SNe data compared to traditional machine classifiers based on random forest or recurrent neural networks.

We present an unsupervised approach using the Multi-Time Attention Network \citep[mTAN][]{mTAN_paper}, which is based on time attention and a generalisation of discrete positional encoding in transformer models to continuous-time. mTAN can handle irregularly spaced values in time and band without any preprocessing for missing values, and while it can be incorporated in any deep neural network, here it is used in an encoder-decoder architecture: it can serve as a feature learner--low-dimensional representations of light curves--with the advantage that these representations directly represent time, or as a generative model. mTAN shares conceptual connections with the temporal modulation approach introduced in \citet{Pimentel_2023} to learn continuous-time representations. However, our approach is structurally simpler: we directly compute attention over the learned time embeddings, whereas they learn a temporal modulation using Fourier decomposition before attention, which offers more interpretability at the cost of additional computation.

This paper focuses on whether mTAN learns physically interpretable latent features from light curves, which is an important precursor to any classification method. These learned features can be particularly valuable for handling the complexity of transient classification arising due to substantial overlap among different classes \citep[e.g., see Figure 19 of][]{Villar_2017} or even a debatable definition of classes themselves, like for SN IIP vs IIL. Another advantage is the added flexibility of combining our approach with any metric-based classification method. Our data consists of light curves based on real detections rather than on simulations to avoid making assumptions about transient population diversity. Only the observed photometric time series is used without auxiliary information about redshift or other properties of hosts. \reviewMoller{This paper is structured as follows. Section \ref{subsubsec:data-collection} describes the data collection and quality filtering, Section \ref{subsec:mtan-approach} provides theoretical details of the mTAN approach, our proposed modification, and training details. Section \ref{sec:results} presents our evaluation of the model, which includes testing generalisation to variable stars and fast-evolving transients absent during training (Section~\ref{subsec:OOD-test}) and an interpretability test to better understand our model's learning mechanism (Section~\ref{subsec:interpret}). Section \ref{sec:discussion} summarises our findings and potential for future work, followed by concluding remarks in Section~\ref{sec:conclusion}.}

\section{Methods}\label{sec:methods}

Section~\ref{subsec:data} describes details of the data set used, including the data collection (Section~\ref{subsubsec:data-collection}) and quality filtering of alerts (Section~\ref{subsubsec:data-quality-filtering}). Section~\ref{subsec:mtan-approach} with theoretical details of the mTAN in Section~\ref{subsubsec:mtan-theory}, details of its integration with the encoder-decoder framework in Section~\ref{subsubsec:encoder-decoder-framework}, details of our methodological modification to the original mTAN in Section~\ref{subsubsec:modifications}, and training details in Section~\ref{subsubsec:training}.

\subsection{Data set preparation}\label{subsec:data}

\subsubsection{Data collection}\label{subsubsec:data-collection}
We poll public alerts from the Zwicky Transient Facility \citep[ZTF;][]{Bellm_2019_ztf} survey from 1st July 2021 to December 31st 2022 (1.5 years) using the Fink broker, a community alert broker designed for time-domain science with Rubin LSST \citep{fink_broker}. ZTF alert data is based on difference imaging for measuring variations in flux and point-spread function photometry. We use the PSF-fit magnitudes ({\it magpsf}) and time of observation ({\it jd}) for $g$ and $r$ bands of alerts, but ignore uncertainties on magnitudes and do not correct for Milky Way or host extinction. Each alert contains a history of photometry measurements for the last 30 days, which allows building light curves using multiple alerts from the same \reviewMoller{object}. The typical cadence of ZTF light curves is around two days.

We restrict our data collection to three types of alerts: (1) those with confirmed Transient Name Server (TNS) classifications of any SNe type: Type Ia and its subtypes (Ia-91T-like, Ia-91bg-like, Ia-CSM, Ia-pec, Iax/02cx-like), Types Ib and its subtypes (Ib-Ca-rich, Ib-pec, Ibn, Ib/c), Ic and its subtypes (Ic-BL, Ic-pec, Icn), Type II and their subclasses (II-pec, IIP, IIL, IIn, IIn-pec, IIb), and superluminous SN (SLSN-I, SLSN-II) (2) those flagged as an early supernova Ia candidate through the classifier within Fink \citep{fink_early_SNIa,Moller2025}, and (3) those with Milliquas \citep[version 8][]{Flesch2023} classifications of the following Active Galactic Nuclei (AGN) types: {\sc QSO} (``Q''-type classes), type-I Seyfert (class ``A''-type classes), BL Lac (class ``B''-type classes), and type-II Seyfert (class ``N''- or ``K''-type classes)--for this purpose, the SIMBAD labels found in Fink are crossmatched with the Milliquas database within 1$\arcsec$\footnote{1621 alerts did not have a corresponding Milliquas match, so we set their class as `Unknown'.}. Alerts in (1) are dominated by type Ia SNe (56\%) and type II SNe (33\%), and alerts in (2) are dominated by QSO (72\%) and type-I Seyferts (20\%).

As our approach is unsupervised in nature, we do not perform any further data curation to handle class imbalance, and we do not use supplementary information such as redshift. However, the class information is useful for analyzing our results, and thus, for simplicity, we focus only on the above three classes instead of all possible types, even though our model can be applied to any type of light curve.

\subsubsection{Quality filtering}\label{subsubsec:data-quality-filtering}
To increase the quality of the sample photometry, only alerts passing the following basic selection cuts are considered: real-bogus quality score > 0.65, number of bad pixels in a $5 \times 5$ pixel region = 0, a Full Width at Half Maximum $\leq$ 5, elongation $\leq$ 1.2, and absolute value of difference magnitude $\leq$ 0.1. Light curves are constructed by time-ordering the set of alerts that pass these criteria. As light curves with very few data points can not constrain the temporal evolution of transients effectively, and thus make evaluation of our model uninformative, we remove those that have fewer than three observations in either the $g$ or the $r$ band. Finally, we remove two other types of light curves: (a) those with object IDs starting with ``ZTF18'' and having one of the TNS SNe type, because Fink received data from November 2019 onwards from ZTF, so most transients with ZTF18 designations would no longer generate alerts by then, except in rare cases; indeed, most of them showed flat light curves, and (b) those whose visual inspection showed multiple detections at the same time--at least ten such cases are found, both arising due to issues with template images used for difference imaging. Apart from these simple selection criteria, we do {\it not} perform any additional model-oriented preprocessing, such as interpolating using GP regression, projecting observations on an even grid, or grouping closely spaced observations in time in order to ensure the evaluation of the model remains transparent for less-than-ideal situations.

The above quality filter criteria result in 15,652 (51,367) light curves corresponding to 437,935 (568,176) alerts, where values inside brackets show statistics before applying the filter criteria. Of the 15,652 light curves, 1389 have one of the SN classes, 10 are early SN Ia candidates, and the rest have one of the AGN classes. Thus, our dataset is biased towards AGN light curves.

\begin{figure}
    \centering
  \includegraphics[width=0.95\linewidth]{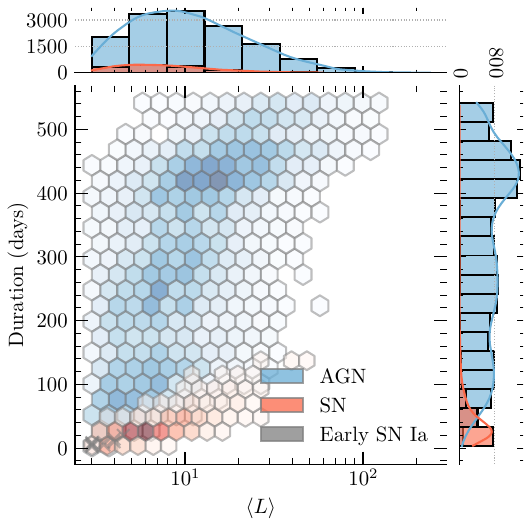}
    \caption{Distribution of duration (difference between the latest and earliest time of observation in days) and no. of observations, averaged across the $g$ and $r$ bands for all polled light curves, separated by three coarse types: SN, early SN Ia candidate, and AGN (see main text for included subtypes). The Early SN Ia candidates are denoted by gray-colored crosses situated near the bottom-left. The 1D histograms on top and bottom axes show the marginal distributions of the average no. of observations and average duration, respectively, where the y-axis shows the counts. A Gaussian kernel density estimate is overlaid on the histograms for visualisation purposes.}
    \label{fig:data-stats}
\end{figure}

Figure~\ref{fig:data-stats} shows a comparison of the number of observations and the duration of light curves. Most SN light curves span less than three months, whereas most AGNs span more than three months, with a significant number that span more than a year, which is expected due to the nature of the variations of SNe and AGNs. Early SN Ia candidates lie at the lower-left end of the plot, around the same location as SN.

In principle, SN and AGN classes can be linearly separated with good accuracy based on a simple cut on the duration (and, optionally, the number of observations), as AGNs tend to have longer durations. However, there are at least two challenges with this simple approach. First, this observed pattern would be unclear had the alerts been polled for only a few consecutive months, i.e., there is a contribution of selection effect due to the time range for which the alerts are polled. Second, this approach would misclassify rare yet astrophysically important transients that do not follow the general trend of that class of transients: for example, SLSN and certain subtypes of SN II are expected to be long-lived but would get misclassified as AGN type. Third, the stochastic nature of AGN variability and the need for a $5\sigma$ detection to trigger a ZTF alert can make AGN light curves appear transient-like when there is flaring activity or even when the variability amplitudes are low, and such events are not unusual in AGNs \citep{Sanchez-Saez_CSAGN}. Thus, we motivate the use of a generalisable and unsupervised approach--representation learning--for characterising light curves in our dataset.

\subsection{Approach}\label{subsec:mtan-approach}

\subsubsection{The Multi-Time Attention Network: mTAN}\label{subsubsec:mtan-theory}

We use the Multi-Time Attention Network (mTAN) approach, which is based on time attention \citep{mTAN_paper}. It learns time embeddings without concatenating them with the input, but by learning to attend to observations at different times using temporal similarity that is calculated from data, that is, without using fixed similarity kernels like those used in traditional GP regression or neural network interpolators with fixed kernel definitions \citep[e.g.,][]{Shukla2019}. The guiding mechanism for attention in mTAN is time embeddings of the query and the key, which are learned during training. Attention in mTAN is directly computed over only observed data points, so it does not require any preprocessing or imputation to handle missing data. Since different light curves may have different numbers of observed data points, computations are parallelized by masking unobserved data points.

The mTAN computation starts with learning continuous-time embeddings through $H$ embedding functions (or heads), $\phi_h(t)$, each of which outputs a representation of size $d_e$. Embedding functions are defined using a combination of periodic terms: $\phi_h(t)[i] = \sin (\omega_{ih} \cdot t + \alpha_{ih})$ for $0 < i < d_e$, and a linear term for the first dimension ($i = 0$): $\omega_{0h} \cdot t + \alpha_{0h}$. These (continuous) time embedding functions can be viewed as a generalisation of the (discrete) positional encodings in transformer networks. These time embedding functions are implemented using a one-layer fully connected layer followed by a sine function non-linearity. The expressivity of the resulting embeddings lies in the fact that time is expressed by learning $d_e$ frequencies and phase shifts of sinusoids for each embedding function.

The mTAN takes in a query time, $t$, and the observed light curve (where the keys are the observed times and the values are the observed magnitudes, for instance), and outputs an embedding vector of dimension $J$ at the query time $t$. Each dimension $j$ ($j < J$) of this mTAN embedding is obtained in broadly two steps. First, continuous-time functions, $\hat{x}_{hd}(t, \mathbf{s})$, are obtained for each dimension of the time series, $d$, and time embedding, $h$, by learning an interpolation kernel:
\begin{equation}
    \hat{x}_{hd}(t, \mathbf{s}) = \sum_{o=1}^{L_d}\kappa_h(t, t_{od})\, x_{od}
\end{equation}\label{eqn:attn-weights}
where $x_{od}$ and $t_{od}$ denote the observed values and times, respectively, for dimension $d$, and $\kappa_h(t, t_{od})$ denotes the attention weights. Here, $\mathbf{s} = (\mathbf{t}, \mathbf{x}) = \{(\mathbf{t_1}, \mathbf{x_1}), ..., (\mathbf{t_N}, \mathbf{x_N})\}$ is the set of observed times, $\mathbf{t_n}$, and observed data (i.e., magnitudes), $\mathbf{x_n}$. The specific form of the kernel, $\kappa_h(t, t_{od})$, is a softmax function with its activation defined by a scaled inner product of the time embeddings of the observed times and the time embeddings of the query time, $t$; thus, the similarity kernel depends only on the time embedding, which is learned from data itself. Second, dimension $j$ of the mTAN embedding is obtained using a linear combination of continuous time functions:
\begin{equation}
    \mathrm{mTAN}(t, \mathbf{s})[j] = \sum_{h=1}^{H} \sum_{d=1}^{D} \hat{x}_{hd}(t, \mathbf{s}) \cdot U_{hdj}
\end{equation}\label{eqn:mtan-eqn}
where $U$ denotes the linear combination weights learned by the model; it is worth noting that $U$ does not depend on the input light curve but is learned globally across the training dataset and is fixed during inference. Such a formulation allows each dimension of the time series (i.e., different bands of a light curve) to use different time embeddings, but also allows information from different dimensions to be shared in the resulting mTAN embedding. Moreover, mTAN embeddings can be calculated for any continuous query time.

The flexibility of mTAN lies in the fact that time embeddings are learnt in a data-driven manner rather than being fixed, which allows for expressing complex kernel functions \citep[][]{mTAN_paper}. Another aspect of flexibility is that the non-linear time embedding introduced by the sine functions means that the query and key time points are not necessarily required to be close together in time to get high attention weights \citep{HetVAE_paper}.

Since the mTAN outputs a continuous function of $t$ (due to the sine non-linearity), it can be useful to discretize its outputs onto a predefined set of reference time points, $\mathbf{r} = [r_1, r_2, ..., r_K]$, which may be uniformly spaced but are not required to be. The mTAN can then be interfaced with neural networks such as recurrent neural networks. The discretized mTAN (or mTAND, for short) then takes in a set of reference time points, $\mathbf{r}$, along with the observed time series and applies mTAN $K$ times, each time using $r_k$ as the query time point ($k = 1$ to $K$); note that the outputs at the $K$ query times are computed parallely instead of sequentially. Thus, mTAND learns representations of dimensions $K \times J$ for each observed light curve.

An important aspect of these learned representations is that the notion of time is included in them, and thus, they encode data's temporal structure in a meaningful manner. This is in contrast to several approaches based on self-supervised learning that use contrastive pretext tasks, for example, \citep[see, e.g.,][for an exception]{Fraikin2023}.

In principle, mTAND can be incorporated into classification architectures and can be followed by any neural network to extract auxiliary information from the regularly spaced latent representations produced by mTAND. In this work, we focus only on light curve representation and follow the mTAN paper by using mTAND in the encoder and the decoder of a Variational Autoencoder (VAE)-like framework and interfacing the outputs of mTAND with a bidirectional gated recurrent unit (GRU). The mTAN framework focuses more on capturing local structure, while the additional GRU component captures global structure in the light curve \citep[][]{mTAN_paper}, which in general has proven more effective than purely attention-based architectures \citep{Tan2025}.

\subsubsection{The encoder-decoder framework}\label{subsubsec:encoder-decoder-framework}

\begin{figure*}
    \centering
  \includegraphics[width=0.9\linewidth]{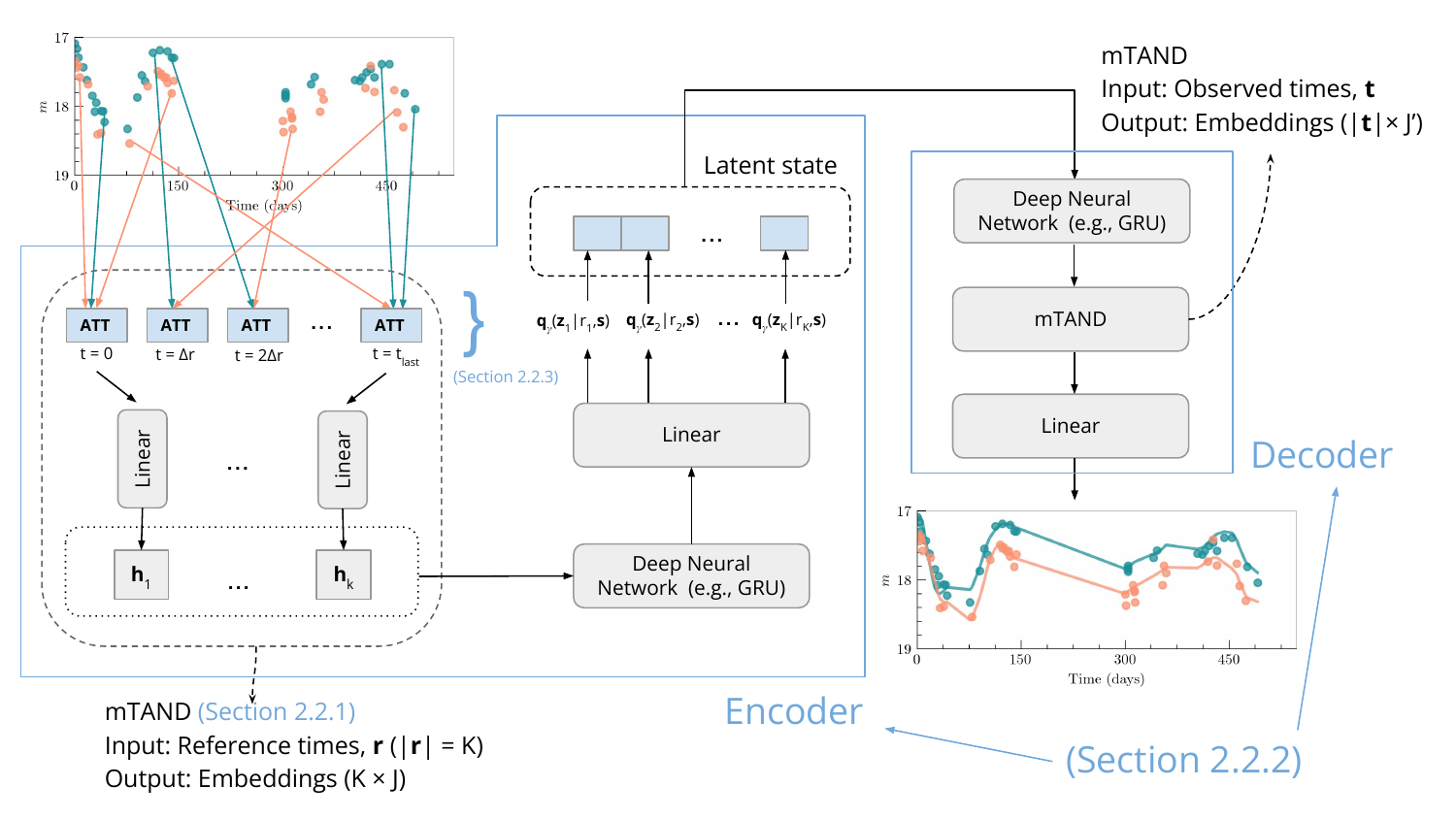}
    \caption{\reviewMoller{Architecture of mTAND and its integration into the encoder-decoder framework. The model takes an irregularly sampled light curve directly as input. {\bf ATT} denotes an attention block implementing Equation~\ref{eqn:attn-weights}, and the linear combination following it (Equation~\ref{eqn:mtan-eqn}) produces the mTAN embedding. Unlike the original mTAN approach, the length, $K$, of the reference times is not fixed but dependent on the input light curve.}}
    \label{fig:mtan-schema}
\end{figure*}

\reviewMoller{Figure~\ref{fig:mtan-schema} provides an overview of the mTAND module and its role within the overall encoder-decoder framework used in this paper.}

First, define a set of latent states at $K$ reference time points: $\mathbf{z} = [\mathbf{z}_1, \mathbf{z}_2, ..., \mathbf{z}_K]$ where each $\mathbf{z}_k$ can be sampled from $P(\mathbf{z}_k) = \mathcal{N}(0, I)$, $I$ being the identity matrix (a multivariate standard normal distribution). The encoder consists of the mTAND module described above, which outputs representations evaluated at the $K$ reference points, followed by a bidirectional GRU with a single recurrent layer. The outputs of the GRU are passed through a set of fully connected layers to return the mean and the variance that are used to construct a Gaussian distribution over $\mathbf{z}$ with diagonal covariance. This distribution, $\mathbf{q}_{\gamma}(\mathbf{z} \given \mathbf{r}, \mathbf{s})$, is the one that is expected to make the observed data likely when sampled from and input to the decoder component.

Here, $N$ is the number of training examples, and $\gamma$ denotes the parameters of the encoder model. Note that each $t_n$ and $x_n$ in the tuple $(\mathbf{t_n}, \mathbf{x_n})$ further comprises of the observed times and data, respectively, for each of the dimensions, $d$, of the light curve; however, we do not explicitly mention it for ease of notation.

The structure of the decoder is the opposite of that of the encoder described above. The latent states sampled from $\mathbf{q}_{\gamma}$ are first passed through a GRU decoder, and the resulting output, along with the observed times as the query (in the encoder, the observed times acted as the key), are passed to the mTAND module to return a sequence of embeddings of length equal to the number of observed time points. These embeddings are passed through a set of fully connected layers, which are then used to parameterize the output distribution, also a Gaussian distribution with diagonal covariance, with mean given by the decoded light curve and a fixed variance.

Let us call this output probability distribution $p_{\theta}(\mathbf{x} \given \mathbf{z}, \mathbf{t})$ where $\theta$ denotes the parameters of the decoder model. During training, the goal is to increase this probability, i.e., gradually improve the model's ability to explain the training data. As is the case for VAEs, in practice, while optimising this probability, we also need to pull $\mathbf{q}_{\gamma}(\mathbf{z} \given \mathbf{r}, \mathbf{s})$ closer to the prior distribution, $p(\mathbf{z})$, which regularises the encoder. Thus, the loss function is as follows:
\begin{align}
\mathcal{L}(\theta, \gamma) = \sum_{n=1}^{N} \frac{1}{\sum_d L_{dn}} \Bigg(
    \mathbb{E}_{q_{\gamma}(\mathbf{z}\mid\mathbf{r}, \mathbf{s}_n)}
    \big[\log p_{\theta}(\mathbf{x}_n\mid\mathbf{z}, \mathbf{t}_n)\big] \nonumber \\
    - \beta_{\text{KL}} \,
    D_{\text{KL}}\!\left(
        q_{\gamma}(\mathbf{z}\mid\mathbf{r}, \mathbf{s}_n)\,\|\,p(\mathbf{z})
    \right)
\Bigg), \\
\intertext{where}
D_{\text{KL}}\!\left(
q_{\gamma}(\mathbf{z}\mid\mathbf{r}, \mathbf{s}_n)\,\|\,p(\mathbf{z})
\right)
= \sum_{k=1}^{K}
D_{\text{KL}}\!\left(
q_{\gamma}(\mathbf{z}_k\mid\mathbf{r}, \mathbf{s}_n)\,\|\,p(\mathbf{z}_i)
\right) \\
\log p_{\theta}(\mathbf{x}_n\mid\mathbf{z}, \mathbf{t}_n)
= \sum_{d=1}^{D} \sum_{l=1}^{L_{dn}}
\log p_{\theta}(x_{ldn}\mid\mathbf{z}, t_{ldn})
\end{align}
\label{eqn:loss-fn}
where $L_{dn}$ is the number of observed data points for dimension $d$ of the $n^{th}$ training example, $D_{\text{KL}}$ denotes the Kullback-Leibler divergence, and $\beta_{\text{KL}}$ is a hyperparameter to control the weight given to the KL term. While the traditional approach of training a VAE maximizes the evidence lower bound, the above loss function normalises\footnote{This normalisation is performed as in the original mTAN paper. However, KL is a regulariser on the latent space, and thus, a better normalisation would be to not normalise the KL divergence term by the number of observations. In our case, $\vert \mathbf{z} \vert$ is not constant, unlike the original mTAN paper, but depends on the light curve duration. Hence, future work would include a better normalisation scheme that appropriately normalises, say, based on duration.} the ELBO using $L_{dn}$ to prevent prioritising light curves with more observations\footnote{Given the imbalance in the total duration of light curves in our data (see Figure~\ref{fig:data-stats}), we tried normalising the contribution of ELBO based on both, the number of observations and light curve duration; however, we did not find improved validation performance during training and so did not use this approach.}. As this loss function does not use information about labels (the type of transient), the training of mTAN in such a setting is unsupervised in nature \citep[see, ][for demonstration of using mTAN for supervised learning using an additional classification module]{mTAN_paper}.

Although the VAE-based framework described above can have other uses during inference (such as generating new samples of light curves using only the decoder and sampling from a standard normal prior), in this paper, we focus on studying the characteristics of encoded representations and interpolation (encoding followed by decoding of light curves).

\subsubsection{Modification to mTAN}\label{subsubsec:modifications}
The original mTAN approach dealt with time series datasets spanning a given duration of time with observations for each datum starting at the same time. Our astronomical transient light curves peak at different times, evolve on widely different time scales, and are sampled differently. As a result, using a predefined fixed number of (equally-spaced) reference points based on the minimum and maximum observed time across training data (as used in the original paper) is not optimal; this is because those would capture the temporal evolution of long-duration light curves (say, those evolving for a year) adequately well but would be too coarse for shorter-duration transients (say, those evolving over only several days).

As light curves of arbitrary durations can be present during training or inference, we modify the calculation of reference times by instead fixing the time interval between adjacent reference times ($\Delta r$ or `temporal resolution') and allowing the number of reference points to vary for each light curve depending on its time span or duration. We use a temporal resolution of two days, which is roughly based on the cadence of ZTF light curves used in this work. The presence of a different number of reference time points for different light curves in a batch is handled by masking for parallel computation. This modification of fixing the time spacing between adjacent reference points rather than the number of reference points still allows a straightforward comparison of the temporal dynamics of any two light curves, irrespective of their duration, because they would have the same reference times. This aspect is crucial as our goal is to characterise light curves but also to compare those evolving over vastly different time scales.

\subsubsection{Training}\label{subsubsec:training}

Our code is a modification of the open-source implementation of mTAN\footnote{\url{https://github.com/reml-lab/mTAN}} and can be found here\footnote{\url{https://github.com/Yash-10/astro-mtan}}. Hyperparameter values are taken from the original mTAN paper. The time embedding dimension $d_e = 128$, the number of embedding functions $H = 1$ (we tried using two heads, but it showed slightly worse validation performance during training), and $D = 4$ (which is twice the number of dimensions of light curves since we have $g$ and $r$ bands; the `two times' arises because the model is used in a VAE-like setup which outputs mean and variance). The hidden size of the GRU encoder, $J = 64$, and that of the GRU decoder is 50, and we use a latent dimension at the output of the encoder (after the GRU) of two, since our light curves have two bands, but we note that the original mTAN paper used much larger values for realistic datasets.

Since we use resolution $\Delta r = 2$ days as mentioned in Section~\ref{subsubsec:modifications}, the number of reference points $K$ is not fixed across the dataset, in general, but caters itself to the duration of individual light curves\footnote{The modifications described in Section~\ref{subsubsec:modifications} subsume the approach of using fixed $K$ when all light curves have equal or almost equal durations in the dataset.}. To handle variable lengths of queries, masking is used to ignore irrelevant queries for a given light curve, similar to attention masking used to ignore unobserved data.

64\% of the light curves are randomly chosen (without using types) for training, 16\% for validation, and the remaining 20\% for testing--out of the 15,652 light curves obtained after quality filtering described in Section~\ref{subsubsec:data-quality-filtering}, this results in 10,016 for training, 2505 for validation, and 3131 for testing. Because our data is biased towards longer duration light curves (see Figure~\ref{fig:data-stats}), a semi-random truncation augmentation is introduced during training to increase the representativeness of shorter duration light curves: a random percentage of observations to sample is selected between 30\% and 70\% followed by randomly selecting a starting observation and then selecting continuous subsequent observations as observed in the original light curve. This augmentation is applied with 50\% probability and the semi-randomness arises because it is only applied to light curves having at least ten observations in each band to prevent extremely sparse light curves.\footnote{Note that random truncation introduces examples with a lesser number of observed data points {\it and} smaller duration, but does not add examples with {\it only} small duration. Although denser temporal sampling is less frequent even when using our augmentation and further increases the `heavy-tailed-ness' of the marginal distribution of the number of observations in Figure~\ref{fig:data-stats}, the normalisation of the loss function by the number of observations (Equation~\ref{eqn:loss-fn}) ensures that the augmentation does not introduce additional data bias while still virtually balancing the presence of short- and long-duration light curves.}

The light curves are normalised using the following normalisation scheme. The observed times in the training, validation, and testing dataset splits are normalised based on the minimum and maximum time across the training set using the min-max scheme ($\dfrac{t - min(t)}{max(t) - min(t)}$); before that, we shift the observed times of all light curves to start at zero, since the absolute times are not relevant for our purposes. The observed (apparent) magnitudes are normalised based on the peak magnitude of the individual light curves ($\dfrac{\mathrm{mag} - min(\mathrm{mag})}{2.5}$), where $min(\mathrm{mag})$ is the peak magnitude across the $g$ and $r$ bands and the scaling factor makes it equivalent to the logarithm of peak flux divided by any observed flux in the light curve. Such a `global' and `local' normalisation of times and magnitudes, respectively, allows the model to focus on light curve shape and evolution trends in magnitude and time, while ignoring apparent brightness. This approach makes the model less susceptible to capturing redshift since it ensures that the model does not learn very different latent representations for transients showing similar temporal evolution, irrespective of their observed magnitudes.

The Adam optimiser is used for optimising the training with a learning rate of $1 \times 10^{-4}$ for both the encoder and the decoder and default momentum parameters: $\beta_1 = 0.9$ and $\beta_2 = 0.99$. A batch size of 8 is used. Training is performed for 300 epochs. The model is validated on a separate validation dataset at each epoch, and the model with the smallest mean squared error between the observed and predicted light curve is saved; this happens at the 298th epoch. To avoid the common problem of vanishing KL term, a monotonic KL annealing schedule is used wherein $\beta_{\text{KL}}$ is set to zero for the first ten epochs and is gradually increased to approach $\beta_{\text{KL}} = 1$ during subsequent epochs. Also, five samples from the distribution $q_{\gamma}(\mathbf{z}|\mathbf{r}, \mathbf{s}_n)$ are used to compute the loss function defined in Section~\ref{subsubsec:encoder-decoder-framework} during training, whereas a single sample is used for validation and testing. The common reparameterization trick that enables optimisation using stochastic gradient methods is used to compute gradients. The fixed variance used for parametrizing the decoder's output distribution is set to $0.01^2$.

The \reviewMoller{model size} of the encoder and decoder networks is roughly 340 KB and 250 KB, respectively, and is thus lightweight.

\section{Results}\label{sec:results}

We evaluate the model on the test set. We first test the (a) decoding of our model to observed light curves (Section~\ref{subsec:interpolation}), and (b) two-dimensional visualisation of latent space and checking for qualitative correlations with basic properties of light curves and non-correlations with non-physical/trivial properties (Sections.~\ref{subsec:latent-space}. 

We then check whether the SN and AGN light curves are represented in different areas in a UMAP and if there are correlations with observed light-curve properties in Section~\ref{subsec:SNAGN-analysis}). In Sections~\ref{subsubsec:sn-analysis} and ~\ref{subsubsec:agn-analysis} we study these correlations more in detail for SN and AGN, respectively.

Following these tests, we visualise the learned attention weights to interpret the model's learning mechanism (Section~\ref{subsec:interpret}), then perform a brief comparison of our model with GP regression (Section~\ref{subsec:gp-reg-compare}), including execution time during inference, and finally showcase application to variable stars and TDEs that the model was \reviewMoller{{\it not}} trained on (Section~\ref{subsec:OOD-test}).

\subsection{Light curve decoding}\label{subsec:interpolation}

The mTAND's decoder models the output Gaussian distribution, $p_{\theta}(\mathbf{x} \given \mathbf{z}, \mathbf{t})$, with its mean given by the decoded light curve, which corresponds to an interpolation over the observed light curve. Here, we study how well these decoded light curves reproduce the observed light curves from the test set. Before discussing the results, we note that because we do not account for observational uncertainties, our interpolations are implicitly overconfident.

\begin{figure*}
    \centering

    \includegraphics[width=0.32\linewidth,keepaspectratio]{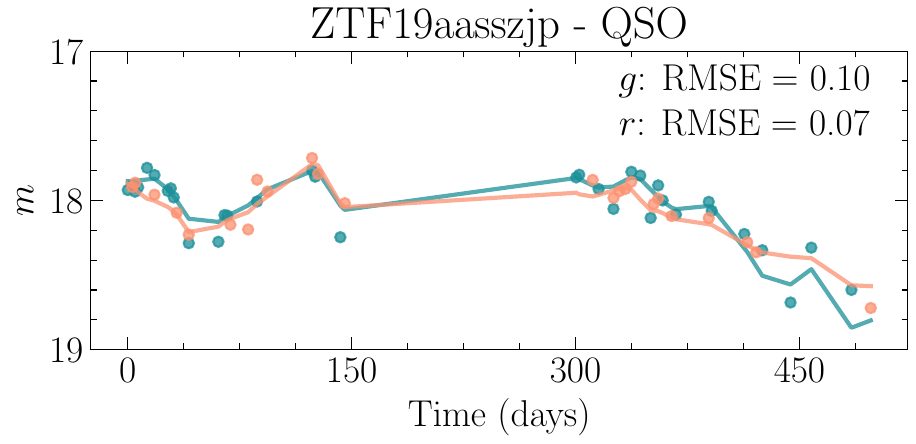}
    \includegraphics[width=0.32\linewidth,keepaspectratio]{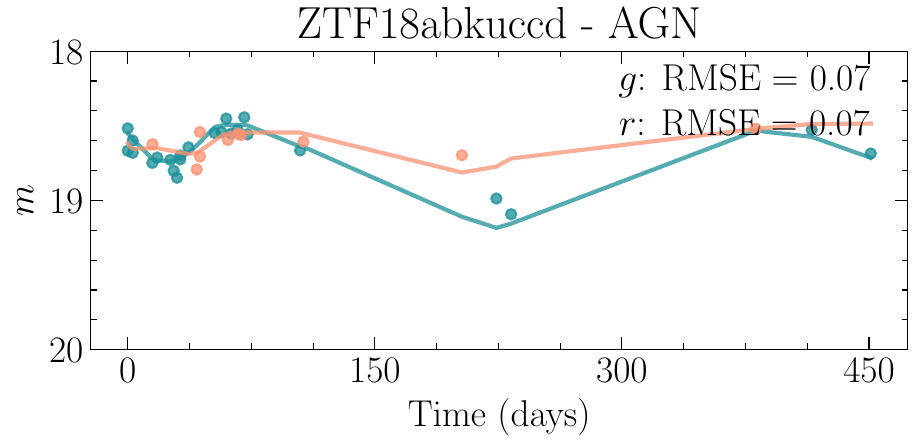}
    \includegraphics[width=0.32\linewidth,keepaspectratio]{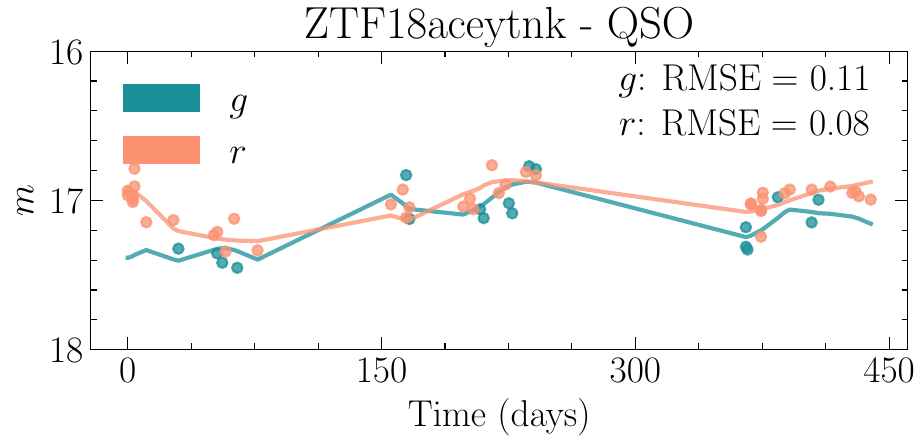}
    \includegraphics[width=0.32\linewidth,keepaspectratio]{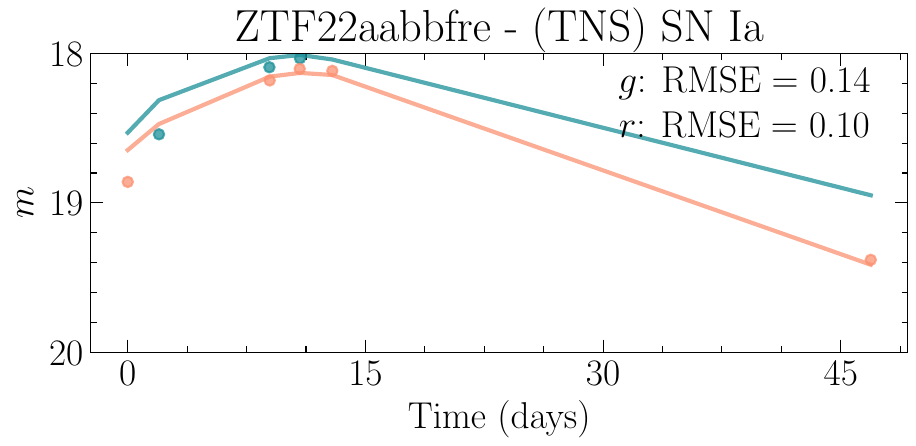}
    \includegraphics[width=0.32\linewidth,keepaspectratio]{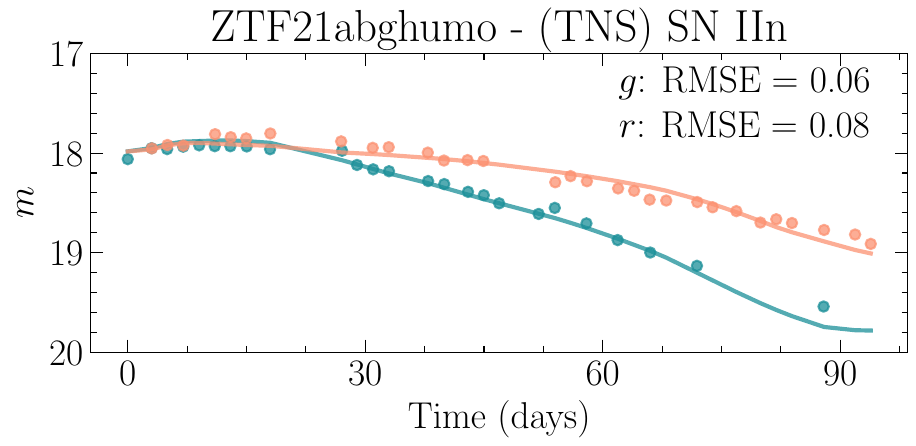}
    \includegraphics[width=0.32\linewidth,keepaspectratio]{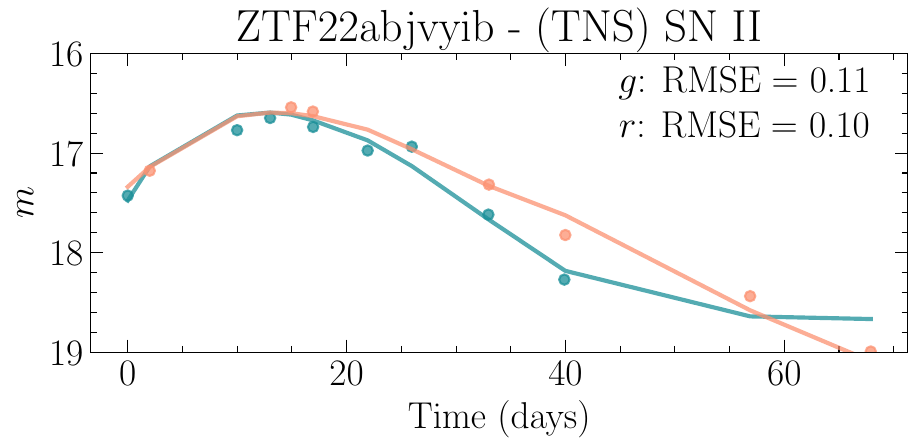}
    \includegraphics[width=0.32\linewidth,keepaspectratio]{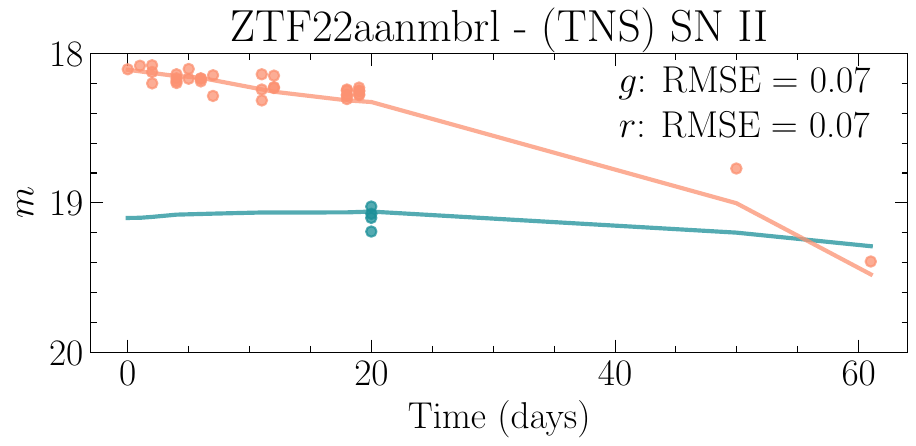}
    \includegraphics[width=0.32\linewidth,keepaspectratio]{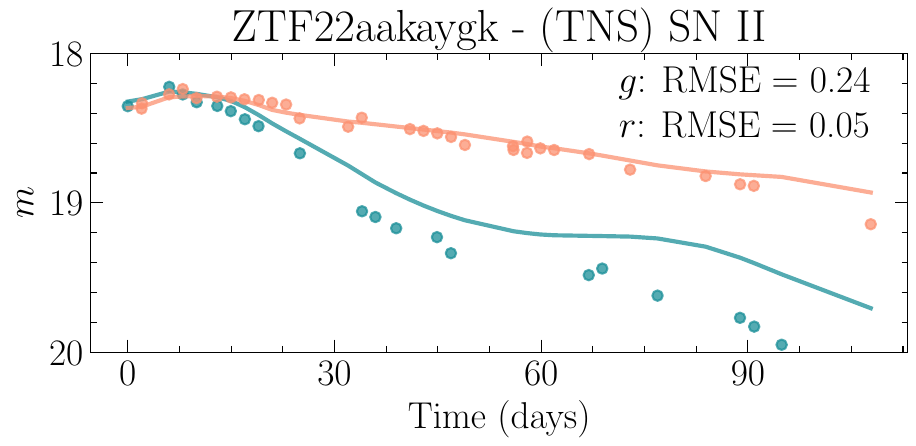}
    \includegraphics[width=0.32\linewidth,keepaspectratio]{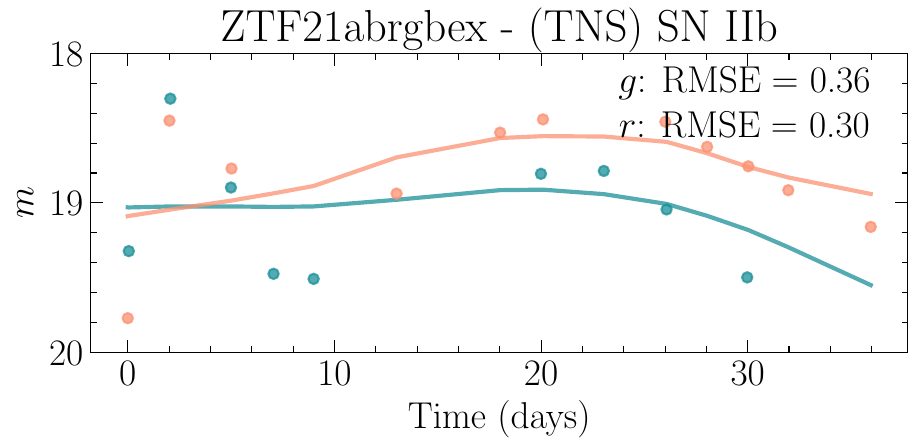}
    \includegraphics[width=0.32\linewidth,keepaspectratio]{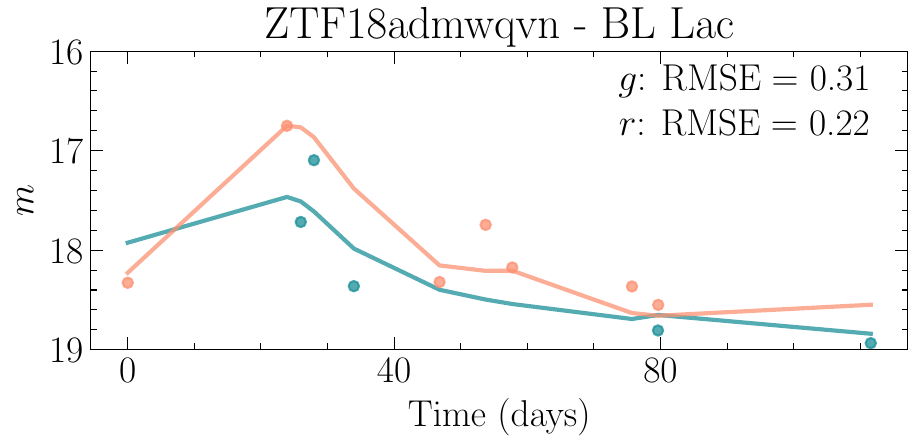}
    \includegraphics[width=0.32\linewidth,keepaspectratio]{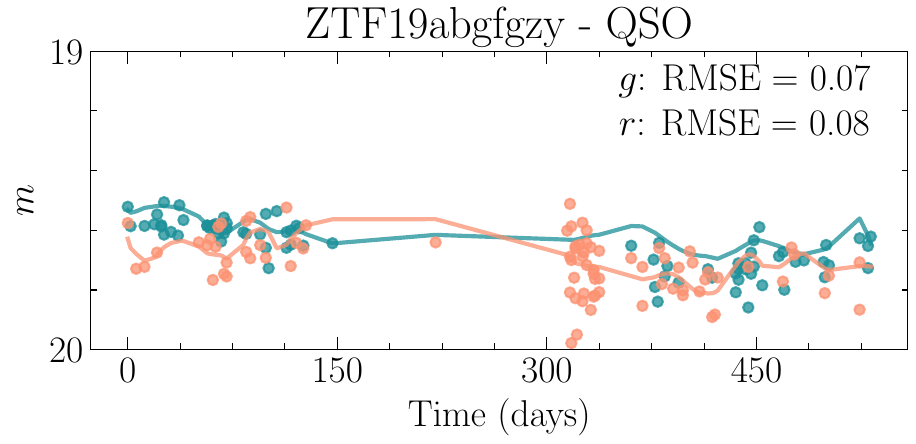}
    
    \caption{Visualisation of a few test-set interpolations by our model. The dots denote observations (uncertainties not shown), and the solid lines denote the interpolations, with the \reviewMoller{blue and red} colors corresponding to the \reviewMoller{$g$ and $r$-bands, respectively}. The root mean square deviation (RMSE) between the observed and predicted magnitudes for each band is shown in each panel. The title of each light curve panel shows the ZTF ID along with its classification from TNS (for SNe) and Milliquas (for AGNs). More interpolation visualisations are shown in Figs.~\ref{fig:sn-light-curves-grid} and ~\ref{fig:agn-light-curves-grid}.}
    \label{fig:interpolation-visualization}
\end{figure*}

Figure~\ref{fig:interpolation-visualization} shows that the mTAN model can accurately reproduce a wide variety of SN and AGN light curves not seen during training. In particular, the model residuals for all light curve observations across the test set have a small bias, mean $\sim$0.01 mag, and a small, in absolute terms, dispersion of $\sim$0.1 mag. Moreover, the interpolations remain good even for sparse light curves with few observations (first row rightmost, second row leftmost), which is not unexpected as the model does not strongly use information about the number of observations (Appendix~\ref{appn:latent-space-extraneous}).

The middle example in the third row shows an example where the general behaviour is not captured well, for which we do not have a clear explanation -- these might indicate problems with the encoding, decoding, or both steps. \reviewMoller{For the last three examples, the general behaviour is captured, but short-time-scale features (e.g., flare around the 300-day mark for ZTF19abgfgzy) are inadequately captured}.

More discussion of subclass-specific light curve features of transients is presented in Section~\ref{subsec:SNAGN-analysis}.

\subsection{Visualisation of latent space}\label{subsec:latent-space}

As mentioned in Section~\ref{subsubsec:encoder-decoder-framework}, the mTAND encoder outputs the mean and variance parameters of the Gaussian distribution learned over latent variables, given a set of reference time points and the input light curve: $\mathbf{q}_{\gamma}(\mathbf{z} \given \mathbf{r}, \mathbf{s})$. To understand the structure of the learned latent space, we visualise the mean latent vector (of length $\lvert \mathbf{r} \rvert$) for each light curve in the test set and discuss qualitative patterns between position in the lower-dimensional representation of the latent space, simple derived properties from light curves, and confirmed classifications from TNS/Milliquas. As $\lvert \mathbf{r} \rvert$ varies for each light curve (see Section~\ref{subsubsec:modifications}), we use zero-padding to have mean latent vectors of the same length and concatenate them across the latent dimension (of size two; see Section~\ref{subsubsec:training}) before projecting them onto a lower-dimensional space.

The projection of the latent space in two dimensions is obtained using Uniform Manifold Approximation and Projection (UMAP), coloured by different light curve features whose choices are motivated below. The implementation of UMAP is taken from \citet{mcinnes2018umap-software}, using the cosine similarity metric and the default hyperparameters, \reviewMoller{including a local neighbourhood size of 15; this controls the tradeoff between local and global manifold structure learned in the encodings, and different values may affect the clustering and its interpretation}

\begin{figure*}
    \centering
    \includegraphics[width=0.32\linewidth,keepaspectratio]{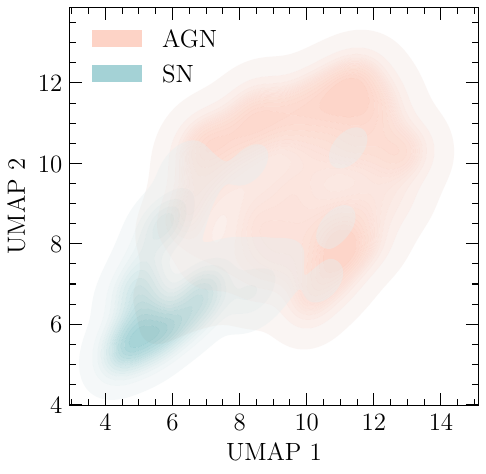}\\
    \includegraphics[width=0.32\linewidth,keepaspectratio]{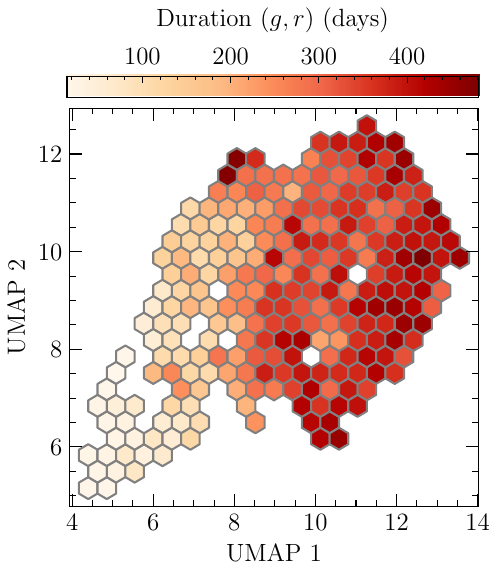}
    \includegraphics[width=0.32\linewidth,keepaspectratio]{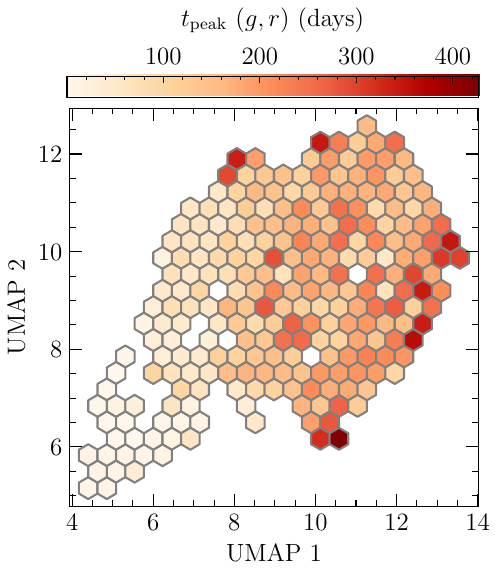}\\
    \includegraphics[width=0.32\linewidth,keepaspectratio]{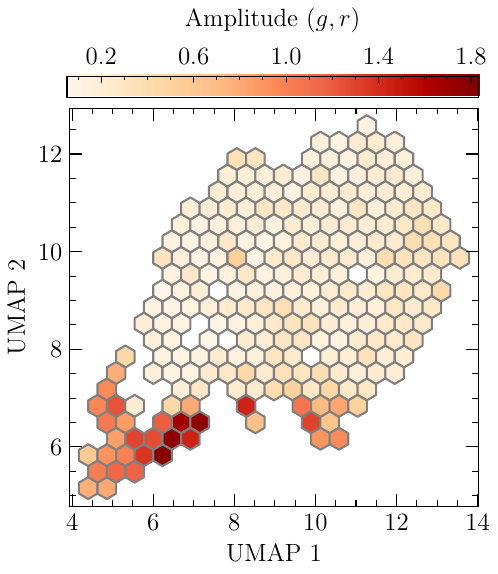}
    \includegraphics[width=0.32\linewidth,keepaspectratio]{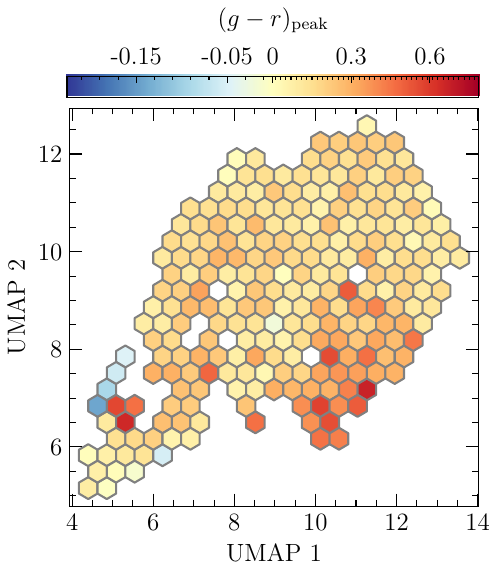}
    \caption{Visualisation of the latent space in two dimensions using UMAP. The top panel shows that the general population of SN (blue) and AGN (red) are separated in the latent space. The next panels show the latent space coloured by different light curve features. Each hexagon bin represents the mean feature of all light curves in that bin, and only bins with at least three light curves are shown. The features used to colour the latent space are: the average of observation time of peak magnitude in the $g$ and $r$ bands with respect to the first observation in the corresponding band ($t_{\mathrm{peak}}\,\,(g, r)$), average of light curve duration in the $g$ and $r$ bands, where duration in each band is based on the time of last and the first observation in that band (Duration $(g, r)$); the same metric is used in Figure~\ref{fig:data-stats}, average amplitude across both bands (A $(g, r)$), and the colour calculated using respective peak magnitudes in $g$ and $r$ bands ($(g - r)_{\mathrm{peak}}$). For this and the remaining figures, wherever applicable, such hand-curated features may be correlated: for example, peak time will necessarily be constrained by duration, and larger duration may increase the probability of observing high-amplitude transients, especially for SNe. \reviewMoller{The patterns with average colour obtained using our model interpolations and standard deviation in colour are similar to the colour at peak and amplitude, respectively.}}
    \label{fig:latent-space-visualization-general}
\end{figure*}

Figure~\ref{fig:latent-space-visualization-general} shows the latent space coloured by features we {\it a priori} expect our model to learn, which are astrophysically important or both. These hand-picked features are: (a) light curve duration, (b) time of peak magnitude since the first observation in the light curve ($t_{\mathrm{peak}}$), (c) difference between the median of top and bottom five percentile magnitudes, and (d) peak colour ($g - r$) \reviewMoller{obtained using respective peak magnitudes}. The heuristics for these choices are as follows: (a) and (b) because the mTAND encodings directly represent time ((a) also because we zero-pad latent vectors corresponding to shorter duration light curves before UMAP projection), (c) because our magnitude normalisation scheme should still preserve the relative magnitude evolution, and (d) because the colour information is preserved since our normalisation uses peak magnitude across $g$ and $r$ bands.

We find that the model learns the overall timescale of the light curve but also when the peak occurs\footnote{In general, it is more appropriate to conclude that the model learns a latent feature that correlates with the feature used for our analysis rather than learning the used feature directly; however, for simplicity in discussion, we would avoid this distinction.}. The small differences in overall trends between the $g$ and $r$ bands for both features are because of minor differences in sampling and data quality. Most light curves have relatively small amplitudes, which is a property of our dataset, and those that have higher amplitudes are visibly separated at the top and top-right. Although not in the standard deviation of magnitudes are very similar, as expected, because both are measures of variability. \reviewMoller{Finally, the plot showing colour based on peak magnitude in the $g$ and $r$ bands suggests that the latent space separates light curves with redder versus bluer colour.} 

Finally, in Appendix~\ref{appn:latent-space-extraneous}, we show that the latent space does not show strong correlations with features our model should not capture, such as peak and mean magnitudes and the total number of observations, or that any observed trends reflect underlying data characteristics. This is desirable for a classifier that must generalise across redshifts and survey depths.

We conclude that the correlations between the positions of light curves in the latent space and our derived features demonstrate that the model has learned multivariate photometric properties of our data.

\begin{figure*}
    \begin{subfigure}{0.32\linewidth}
        \includegraphics[width=\linewidth,keepaspectratio]{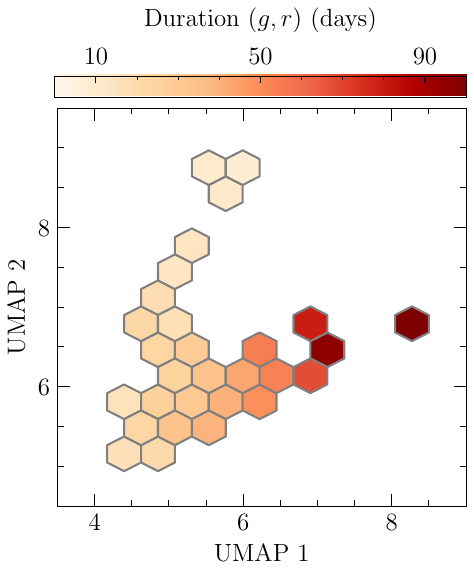}
        \caption{}\label{fig:latent-space-only-sn-features-dur}
    \end{subfigure}
    ~
    \begin{subfigure}{0.32\linewidth}
        \includegraphics[width=\linewidth,keepaspectratio]{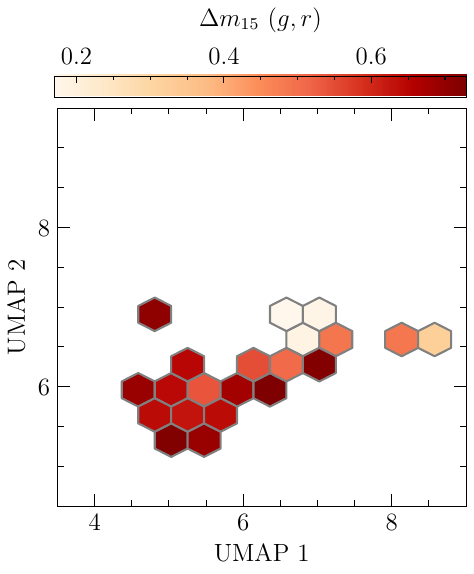}
        \caption{}\label{fig:latent-space-only-sn-features-declinerate}
    \end{subfigure}
    ~
    \begin{subfigure}{0.32\linewidth}
        \includegraphics[width=\linewidth,keepaspectratio]{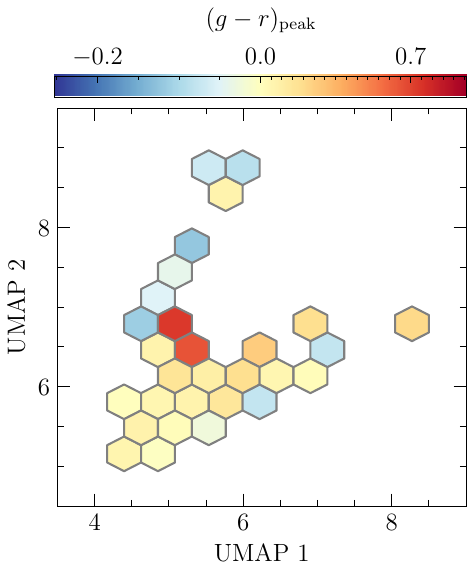}
        \caption{}\label{fig:latent-space-only-sn-features-colorpeak}
    \end{subfigure}
    \caption{Latent space visualisation for only SN light curves, including all its subtypes. Only bins with at least three data points are shown, and thus some outliers from Figure~\ref{fig:latent-space-only-sn-class-colored} may not be included. The features used are: the mean of $r$- and $g$-band (a) duration, (b) decline rate, defined as the decrease in magnitude 15 days after peak time ($\Delta m_{15}$) in either band, and (c) $g - r$ colour at peak. As our mTAND model's interpolations fill unobserved dimensions of light curves in either band, but are not continuous functions of time, the magnitude 15 days post-peak is calculated using linear interpolation on our individual band interpolations. The panels showing the decline rate and peak colour do not show clear patterns but suggest weak trends. For example, bluer peak colors tend to be separated from mildly red/no-colour light curves, and slow decline rates in both bands seem to be separated from faster decline rates in at least one band. The trends in these plots are used in Section~\ref{subsubsec:sn-analysis} to explain the positions of SN light curves from Figure~\ref{fig:latent-space-only-sn-class-colored}. Finally, while these plots show the mean features in each bin, in Appendix~\ref{appn:latent-space-std-features} we show their $1\sigma$ standard deviation, which, when compared with the mean features shown here, suggests that the variations in each bin are not huge.}\label{fig:latent-space-only-sn-features}
\end{figure*}

\begin{figure*}
    \begin{subfigure}{0.75\linewidth}
        \includegraphics[width=\linewidth,keepaspectratio]{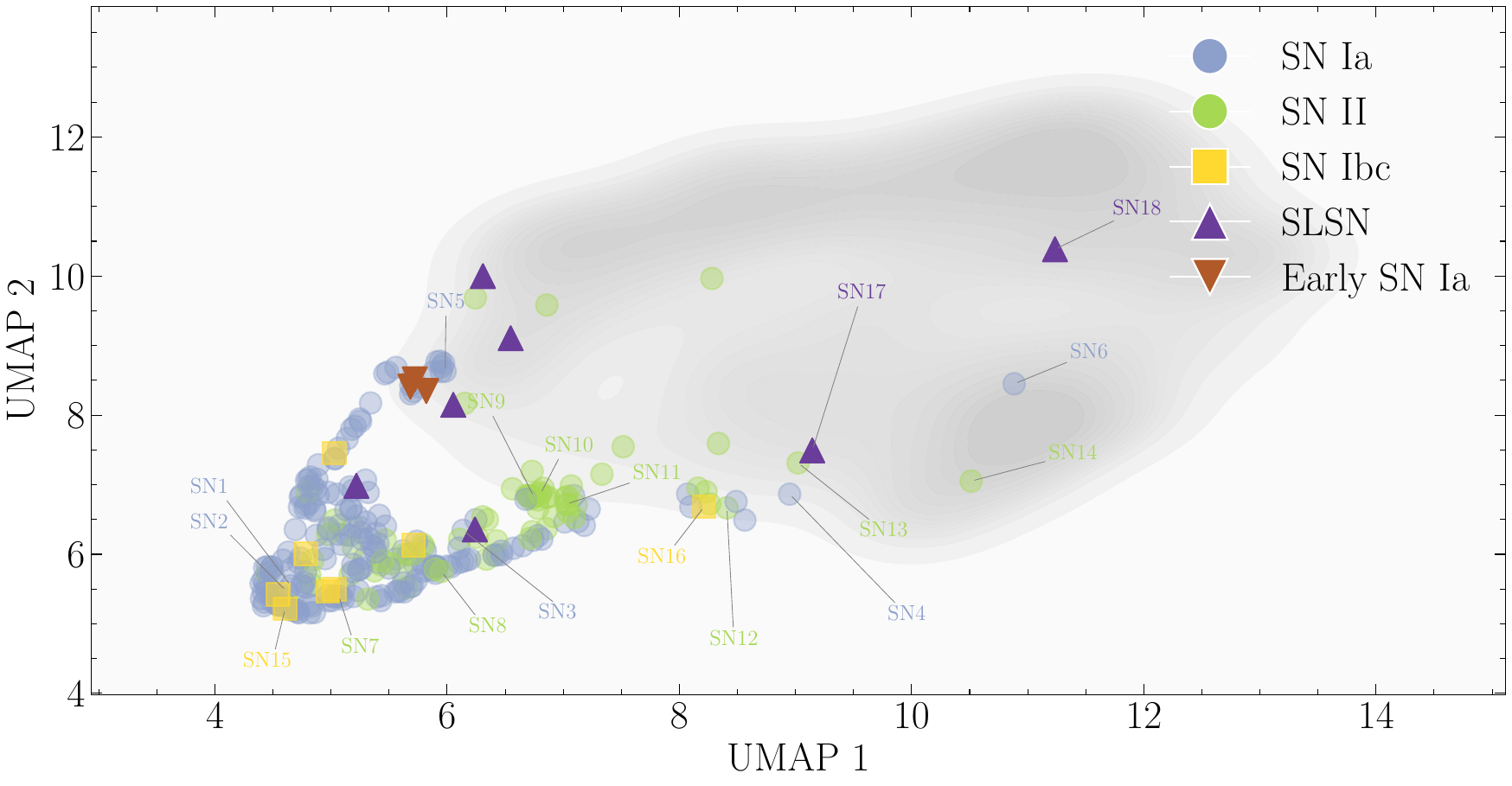}
        \caption{}
        \label{fig:latent-space-only-sn-class-colored}
    \end{subfigure}
    \begin{subfigure}{0.96\linewidth}
        \includegraphics[width=0.32\linewidth,keepaspectratio]{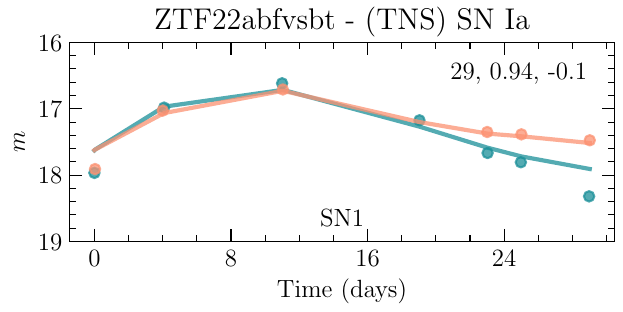}
        \includegraphics[width=0.32\linewidth,keepaspectratio]{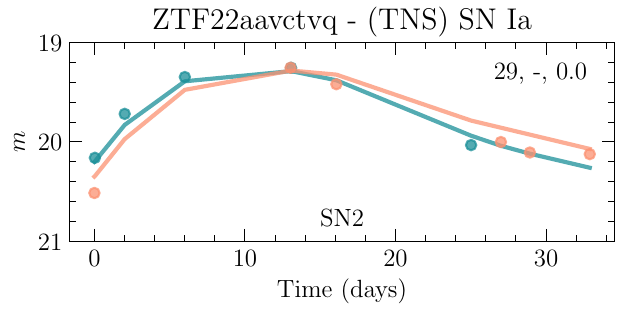}
        \includegraphics[width=0.32\linewidth,keepaspectratio]{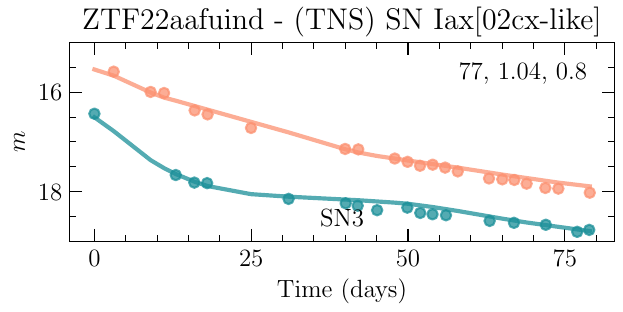}
        \includegraphics[width=0.32\linewidth,keepaspectratio]{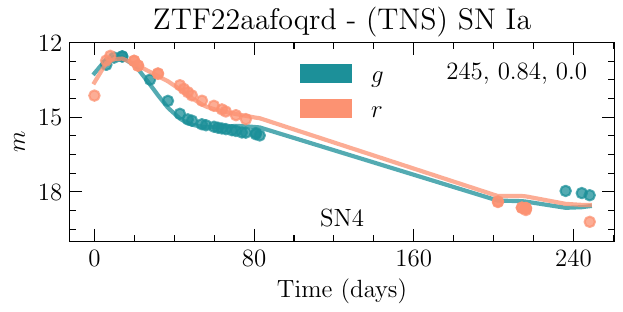}
        \includegraphics[width=0.32\linewidth,keepaspectratio]{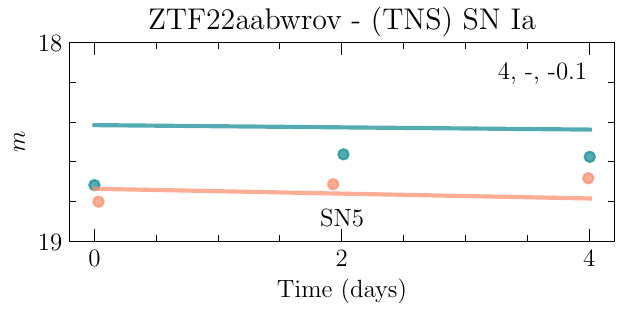}
        \includegraphics[width=0.32\linewidth,keepaspectratio]{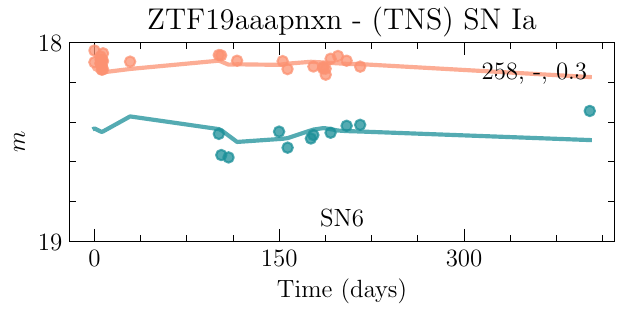}
        \includegraphics[width=0.32\linewidth,keepaspectratio]{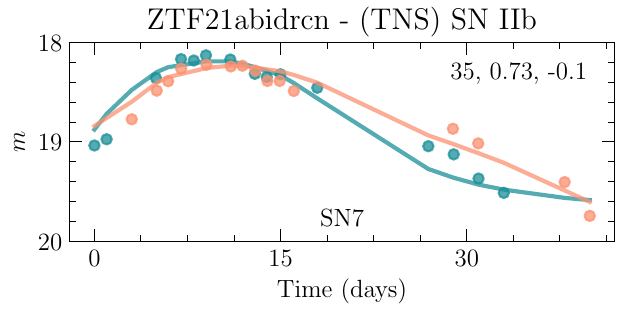}
        \includegraphics[width=0.32\linewidth,keepaspectratio]{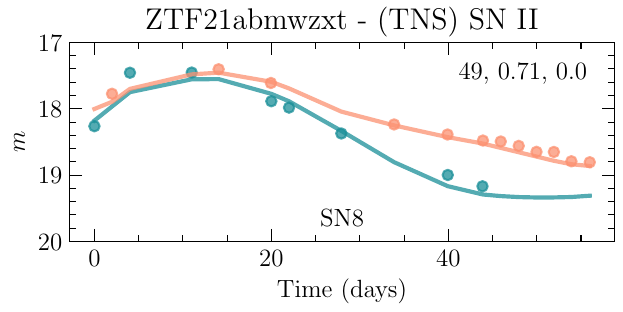}
        \includegraphics[width=0.32\linewidth,keepaspectratio]{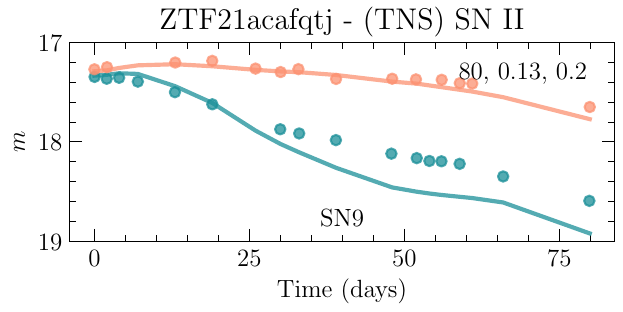}
        \includegraphics[width=0.32\linewidth,keepaspectratio]{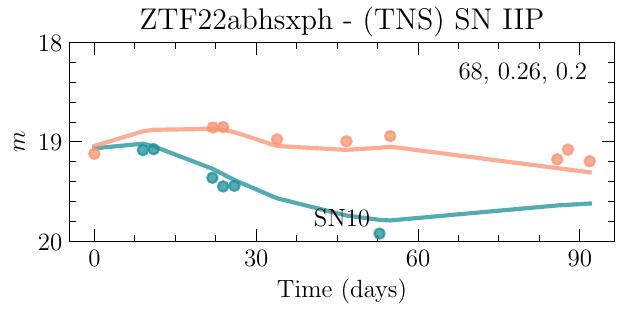}
        \centering
        \includegraphics[width=0.32\linewidth,keepaspectratio]{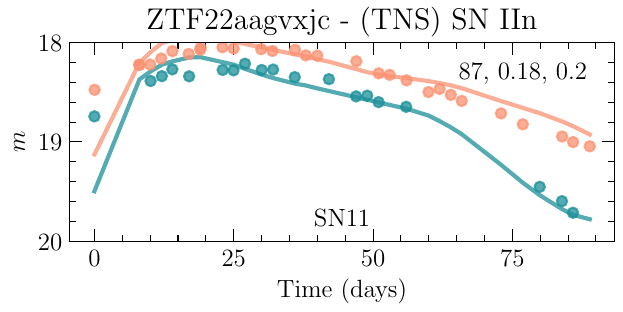}
        \includegraphics[width=0.32\linewidth,keepaspectratio]{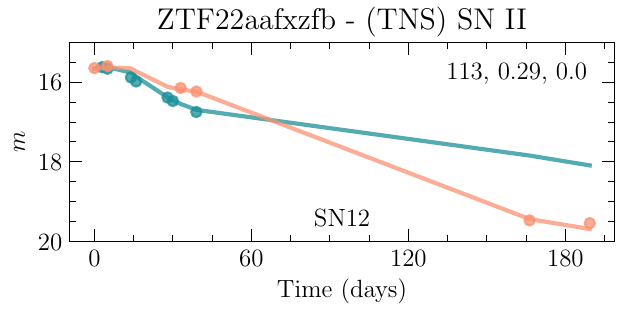}
        \includegraphics[width=0.32\linewidth,keepaspectratio]{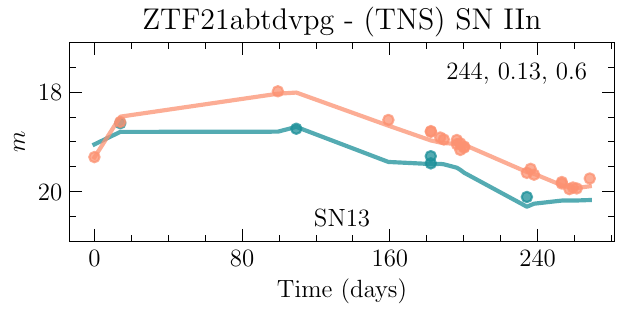}
        \includegraphics[width=0.32\linewidth,keepaspectratio]{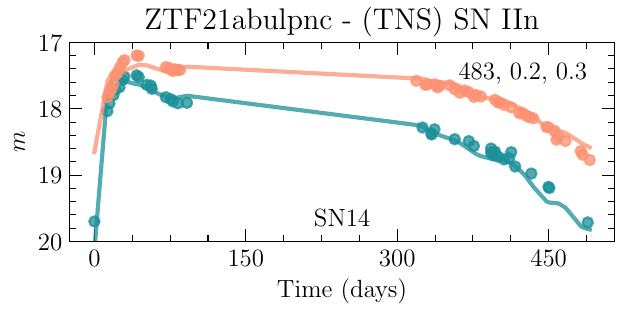}
        \includegraphics[width=0.32\linewidth,keepaspectratio]{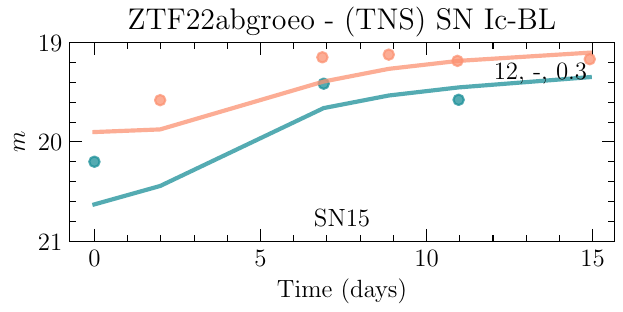}
    \end{subfigure}
\end{figure*}
\begin{figure*}
\ContinuedFloat
    \begin{subfigure}{0.96\linewidth}
        \includegraphics[width=0.32\linewidth,keepaspectratio]{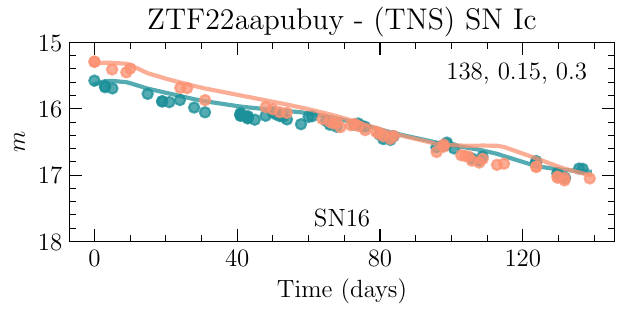}
        \includegraphics[width=0.32\linewidth,keepaspectratio]{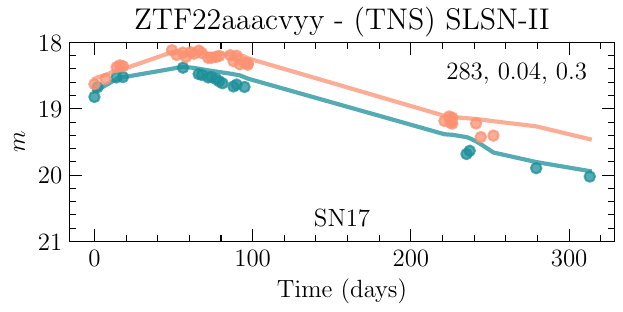}
        \includegraphics[width=0.32\linewidth,keepaspectratio]{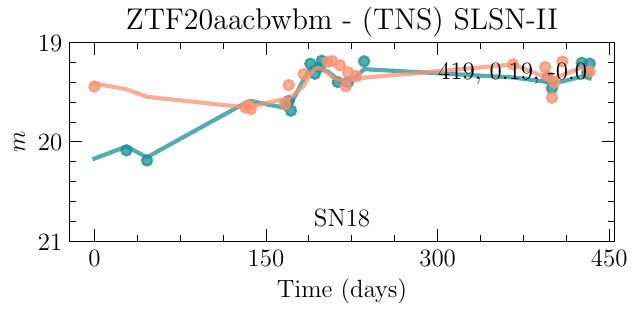}
    \caption{}
    \label{fig:sn-light-curves-grid}
    \end{subfigure}
        \caption{(a) Visualisation of the latent space like in Figure~\ref{fig:latent-space-visualization-general} but colored based on five SN subtypes based on TNS classifications (in decreasing order of prevalence in our dataset): (1) SN Ia and subtypes, (2) SN II and subtypes, (3) SN Ib and Ic; we also include the `SN I' TNS class, (4) SLSN and subtypes, and (5) Early SN Ia candidates (see Section~\ref{subsubsec:data-collection} for specific subtypes included). The gray contour shows the distribution of AGN data points, including all their subtypes. \reviewMoller{(b) The panels below show a few light curves of different SN subtypes located at different locations in the latent space.} The title of each light curve panel shows the ZTF ID along with its spectroscopic classification from TNS. Also shown at the top-right of each light curve are the duration (in days), decline rate ($\Delta m_{15}$), both averaged across $g$ and $r$ bands, and colour at peak, in that order. These are also used in Figure~\ref{fig:latent-space-only-sn-features} and ~\ref{fig:latent-space-only-sn-features-remaining}. A hyphen for the decline rate indicates it could not be calculated because at least one band's light curve terminated before 15 days after peak (see Figure~\ref{fig:latent-space-only-sn-features} caption). 
        }\label{fig:latent-space-sn}
\end{figure*}

\begin{figure*}
    \centering
    \begin{subfigure}{0.75\linewidth}
        \includegraphics[width=\linewidth,keepaspectratio]{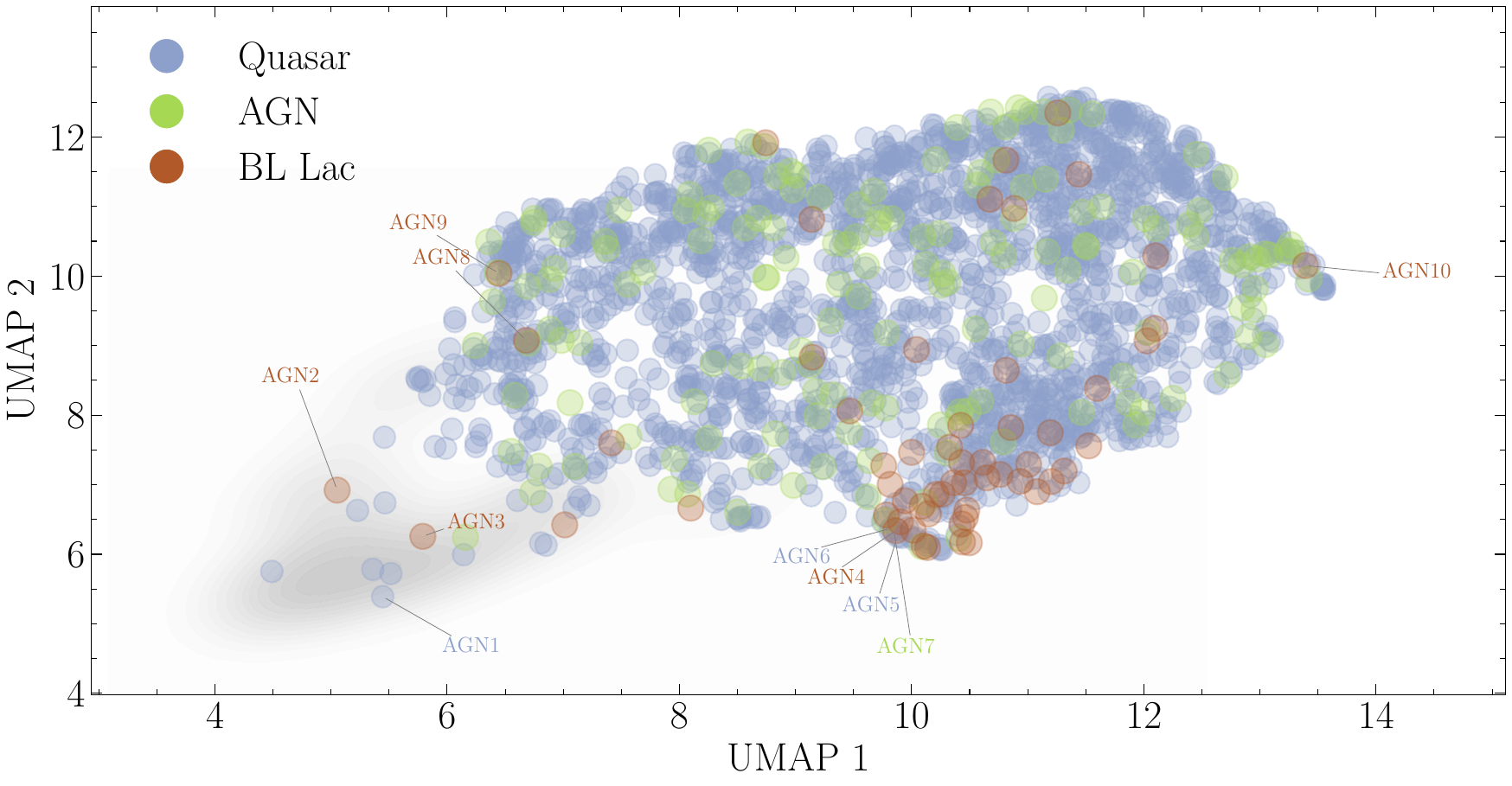}
        \caption{}\label{fig:latent-space-only-agn-class-colored}
    \end{subfigure}
    \begin{subfigure}{0.96\linewidth}
        \includegraphics[width=0.32\linewidth,keepaspectratio]{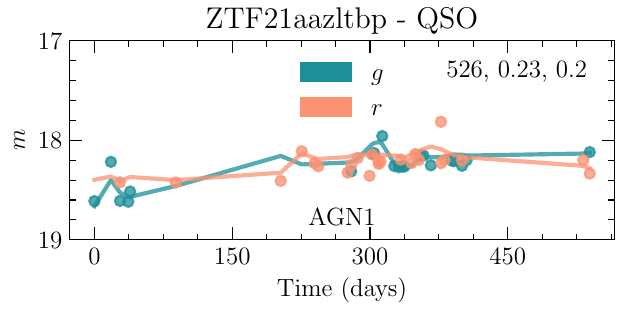}
        \includegraphics[width=0.32\linewidth,keepaspectratio]{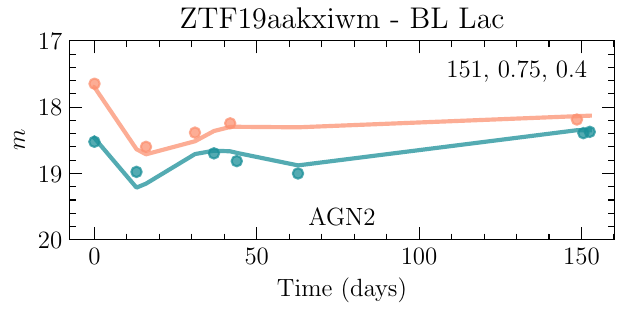}
        \includegraphics[width=0.32\linewidth,keepaspectratio]{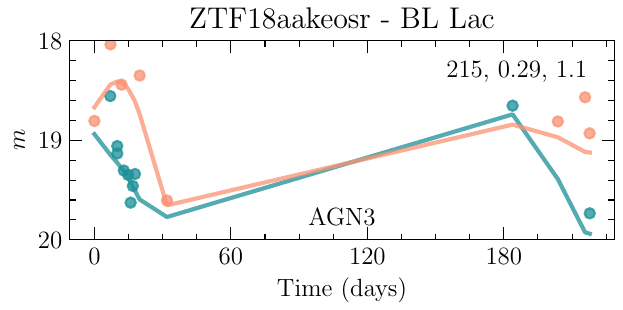}
        \includegraphics[width=0.32\linewidth,keepaspectratio]{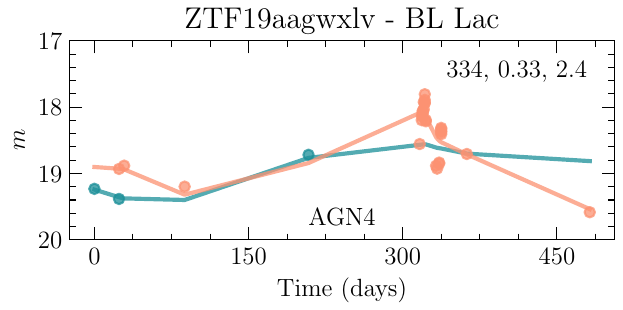}
        \includegraphics[width=0.32\linewidth,keepaspectratio]{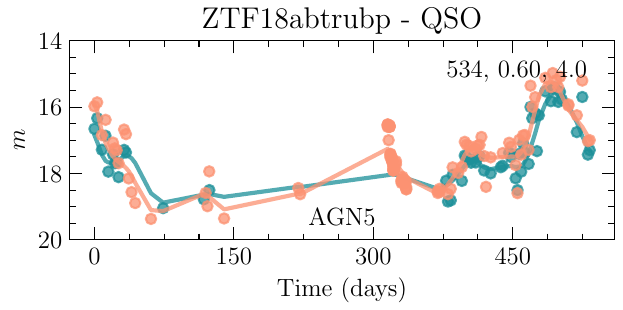}
      \includegraphics[width=0.32\linewidth,keepaspectratio]{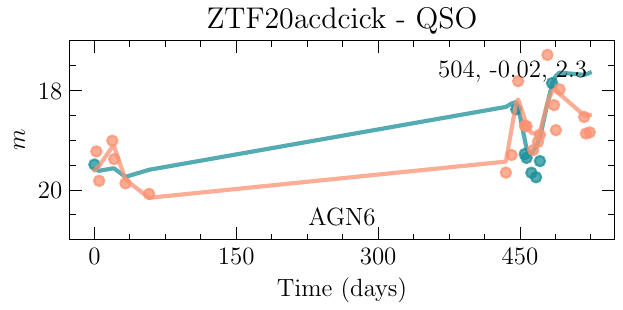}
      \includegraphics[width=0.32\linewidth,keepaspectratio]{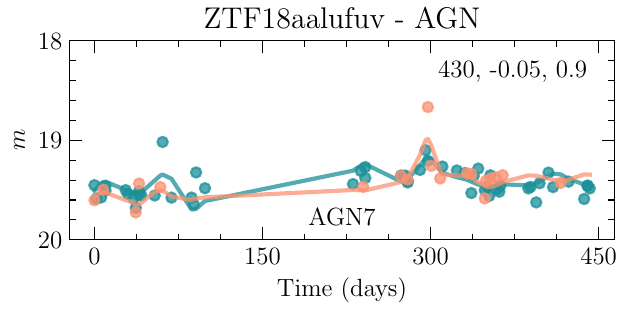}
      \includegraphics[width=0.32\linewidth,keepaspectratio]{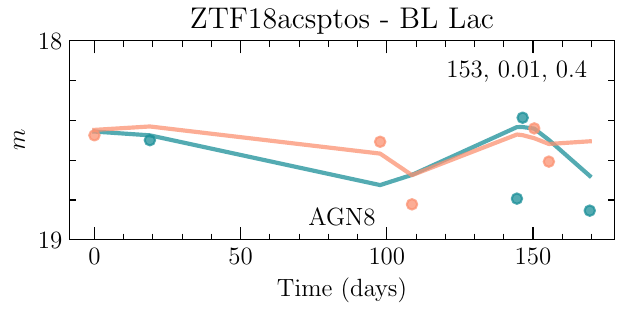}
      \includegraphics[width=0.32\linewidth,keepaspectratio]{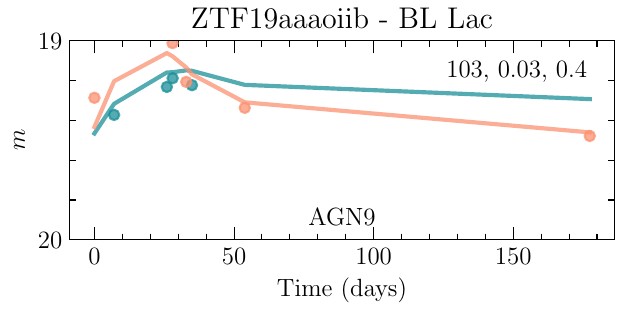}
      \centering
      \includegraphics[width=0.32\linewidth,keepaspectratio]{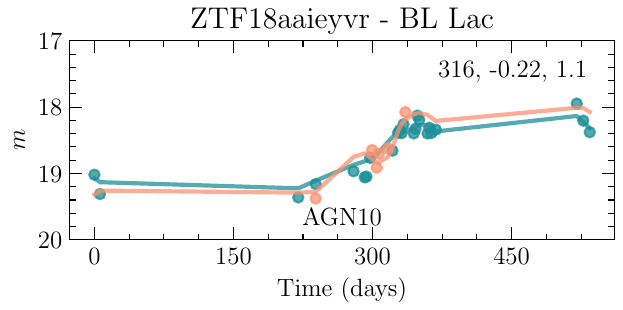}
    \caption{}
    \label{fig:agn-light-curves-grid}
    \end{subfigure}
    \caption{Visualisation of the latent space like in Figure~\ref{fig:latent-space-visualization-general} but colored based on the following AGN subtypes based on Milliquas classifications (in decreasing order of prevalence in our dataset): (1) QSO, (2) Type I Seyfert (or AGN), and (3) BL Lac. \reviewMoller{As mentioned in the text (Section~\ref{subsubsec:agn-analysis}), we remove <3 month AGNs or those with no Milliquas match from this plot; this results in the removal of all four type II Seyferts in our data}. The gray contour shows the distribution of SN data points, including all their subtypes. \reviewMoller{The panels below show a few light curves of different AGN subtypes located at different locations in the latent space}. The title of each light curve panel shows the ZTF ID along with its Milliquas classification. Also shown at the top-right of each light curve are the duration (in days), mean color, and $\sigma/\mu$ amplitude, in that order, where the two metrics apart from mean colour are averaged across $g$ and $r$ bands. These are also used in Figure~\ref{fig:latent-space-only-agn-features}. 
    }\label{fig:latent-space-agn}
\end{figure*}

\subsection{SN and AGN latent space}\label{subsec:SNAGN-analysis}

In the next two subsections, we provide an interpretation of why a light curve representation is mapped to a specific UMAP region. To achieve this, we colour the latent space projection using TNS and Milliquas classifications and compare the known photometric behaviour of supernova and AGN subtypes with what the model has learned.

Before discussing the detailed results, we point out that the UMAP visualisation of latent space separates the SN and AGN classes, as shown in Figure~\ref{fig:latent-space-visualization-general}. This is despite the imbalance in the number of examples in our data and the use of partial light curves. This (unsupervised) separation of SN and AGN is because of differences in their general photometric properties shown by the general analysis in the previous subsection. In particular, the region of the projected latent space where SNe lie is occupied by light curves with an overall shorter duration and slightly larger amplitudes than regions occupied by most AGNs.

\subsection{Supernovae}\label{subsubsec:sn-analysis}

Figure~\ref{fig:latent-space-only-sn-features} shows that duration is the primary organising axis of the SN latent space. Peak colour and decline rate show weaker but perceptible correlations. We use these trends as a reference frame for the rest of this section. 

\subsubsection{Subtype overlaps and their physical origin}

Different SN subtypes overlap in the latent space for three physically distinct reasons, which we discuss in turn.

\textit{Similar photometric timescales (SN~Ia and SN~Ibc).} SNe~Ia and Ibc share similar decline timescales and/or duration, causing their latent regions to overlap substantially. SNe Ibc lying near SN1 and SN2 (e.g., SN15) occupy these regions for this reason. Since the latent space does not strongly encode peak colour, the tendency of SN~Ibc to be redder at peak does not separate them from SN~Ia here. SN1 and SN2 lie close despite a roughly 2.5~mag difference in peak brightness, which confirms that the model does not rely on observed magnitude.

\textit{Partial light curves missing discriminating features (SN~II near SN~Ia)} SN~II light curves that overlap with SN~Ia typically lack plateau coverage in the polled window. SN7 (IIb) and SN8 decline rapidly without a visible plateau, producing SN~Ia-like timescales and latent positions. By contrast, SN9 (showing an $r$-band plateau), SN10 (IIP), and SN11 (a slow-declining IIn with plateau followed by faster decline) occupy a distinct region defined by long duration and slow decline rate, clearly separated from most SN~Ia. A few SN~Ia with slow, plateau-like $r$-band declines lie nearer to these SN~II than to the typical SN~Ia. The latent space, therefore, does distinguish plateau-like evolution when sufficient coverage is present in the polled light curve. We do not use peak times in this part of the discussion since plateaus near peaks can produce misleading peak time estimates.

\textit{Absent luminosity information (SLSN).} Without redshift or absolute magnitude, SLSN cannot be separated from their photometric counterparts at similar apparent timescales. Most SLSN occupy relatively empty regions near the boundaries between the SN and AGN subpopulations. SN17 (SLSN-II) and SN13 (SN~IIn) lie close together, which is physically understandable since SLSN-II are considered a (more luminous) subtype of IIn \citep{GalYam2012}. SN18 lies near AGNs: inspection of its full light curve reveals post-peak observations containing a broad secondary feature with low variability, which is photometrically AGN-like at that phase. We find no overlaps between our SLSN and those present in the ALeRCE \citep[][]{Sanchez_2021} or Superphot+ predictions of \citet{spp_zenodo}, which precludes direct comparison.

\textit{Early SN~Ia candidates.} The three early SN~Ia candidates capture only the very first days of SN~Ia evolution, showing negligible magnitude variation and very small durations (see also SN5). They cluster away from the main SN~Ia region because the model associates their near-featureless, low-amplitude light curves with a distinct corner of the latent space. As more observations of such candidates accumulate, their latent positions would be expected to migrate towards the main SN~Ia population.

\subsubsection{Outlier positions as diagnostics}

A handful of objects lie far from their class's main latent region, and their positions are instructive. SN4 (Ia) has an anomalously long eight-month decay with a very large amplitude, untypical of SN~Ia in this sample, which places it in a sparsely populated outlier region not captured in Figure~\ref{fig:latent-space-only-sn-features}. SN6 contains noise-dominated late-time observations after the explosion, as we found by inspecting its full light curve, and expectedly drifts toward AGN. SN14 (SN IIn) lies near high-amplitude AGNs (Figure~\ref{fig:latent-space-only-agn-features}). SN12 (SN~II) sits in an outlier region characterised by very long duration, low-to-moderate decline rate, and mildly red peak colour, qualitatively matching SN16 (SN Ic).

SN3 represents a peculiar SN~Ia (SN~Iax) and its nearest neighbours in the latent space contain SN~Ia, SN~II, and SLSN subtypes. It was classified as SN~II by the ALeRCE supernova classifier \citep{Sanchez_2021}, while Superphot+ \citep{deSoto_2024} assigned the highest probability to SN~Ibc, followed by SN~Ia, with a low probability to SN~II \citep[predictions from][]{spp_zenodo}. Our unsupervised placement is therefore intermediate between these approaches. Although not shown, two more SN~II in our test set overlap with objects present in the ALeRCE and Superphot+ catalogues. The first, located in the cluster containing SN9 and SN10 and surrounded mainly by SN~II, was classified as SN~II by both approaches; our expected prediction agrees with both approaches. The second, surrounded by a mix of SN~Ia, SN~II, one SN~Ibc, and one SLSN, was classified as SN~II by ALeRCE, while Superphot+ gave a low-confidence SN~Ia classification with SN~II as the next most probable class; our expected prediction aligns more closely with Superphot+ in this case. Direct comparisons with other approaches should be interpreted with caution because our polled light curves, and thus the features available to each model, may differ from those used in other studies. We find no overlaps between our SN~Ibc sample and the ALeRCE/Superphot+ catalogues, which precludes comparison for that subtype.

\subsection{AGN}\label{subsubsec:agn-analysis}

\begin{figure*}
    \begin{subfigure}{0.32\linewidth}
        \includegraphics[width=\linewidth,keepaspectratio]{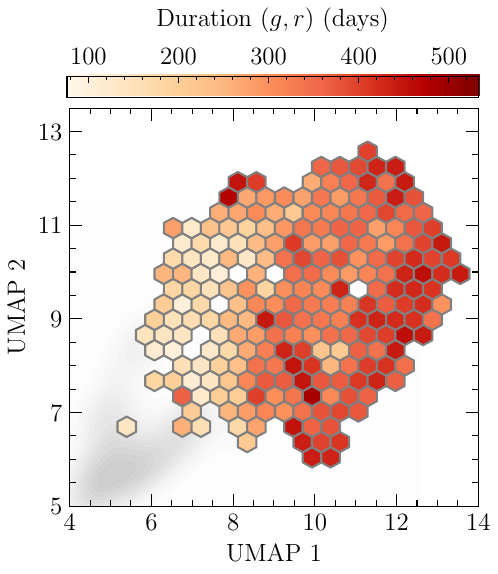}
        \caption{}\label{fig:latent-space-only-agn-features-dur}
    \end{subfigure}
    ~
    \begin{subfigure}{0.32\linewidth}
    	\includegraphics[width=\linewidth,keepaspectratio]{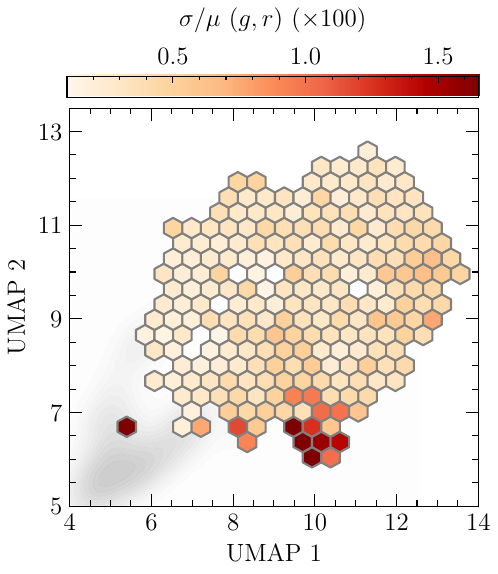}
    	\caption{}\label{fig:latent-space-only-agn-features-meanvariance}
    \end{subfigure}
    ~
    \begin{subfigure}{0.32\linewidth}
        \includegraphics[width=\linewidth,keepaspectratio]{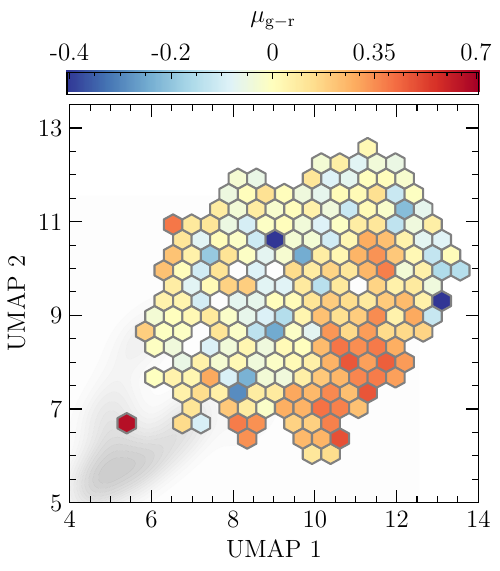}
        \caption{}\label{fig:latent-space-only-agn-features-meancolor}
    \end{subfigure}
    ~
    \begin{subfigure}{0.32\linewidth}
        \includegraphics[width=\linewidth,keepaspectratio]{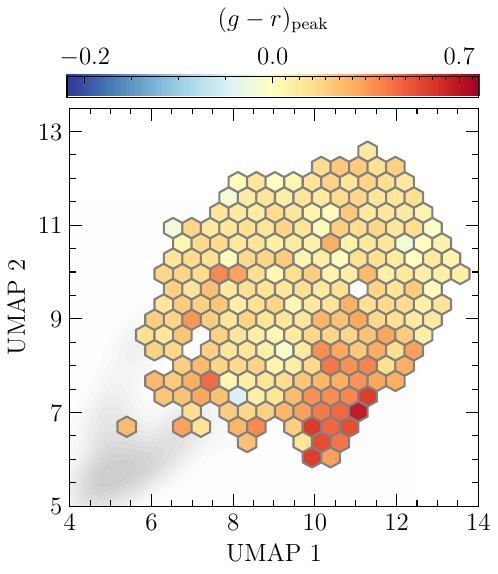}
        \caption{}\label{fig:latent-space-only-agn-features-peakcolor}
    \end{subfigure}
    ~
    \begin{subfigure}{0.32\linewidth}
        \includegraphics[width=\linewidth,keepaspectratio]{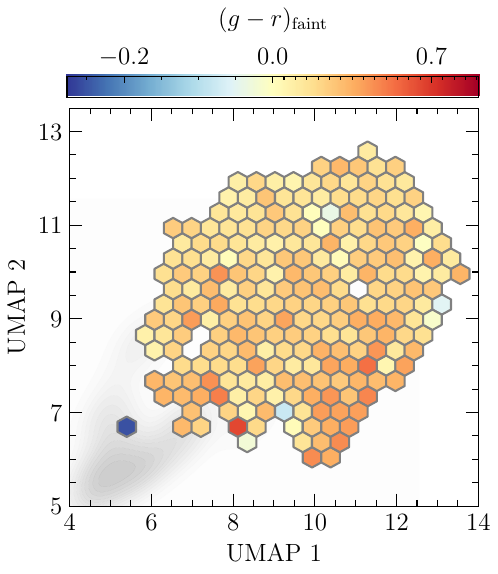}
        \caption{}\label{fig:latent-space-only-agn-features-faintestcolor}
    \end{subfigure}
    ~
    \begin{subfigure}{0.32\linewidth}
        \includegraphics[width=\linewidth,keepaspectratio]{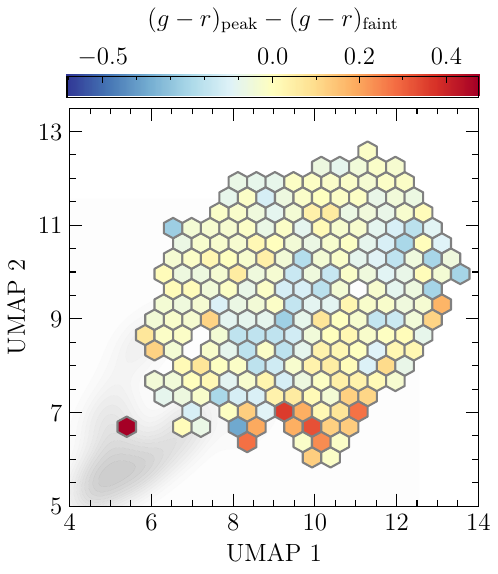}
        \caption{}\label{fig:latent-space-only-agn-features-peakminusfaintestcolor}
    \end{subfigure}
    \caption{{\it Figure description}: Latent space visualisation for only AGN light curves, including all its subtypes. \reviewMoller{As mentioned in the text, AGNs with duration <3 months are excluded.} Only bins with at least three data points are shown. The features used are: the mean of $r$- and $g$-band (a) duration, (b) ratio of standard deviation and mean magnitude, a common variability measure \citep[see, e.g.,][]{Kim2011}, expressed as a percentage, (c) mean color, \reviewMoller{(d) peak color, (e) colour using respective faintest magnitudes, and (f) difference of (d) and (e)}. Since inspection of visual correlations is sensitive to value ranges, we have set colorbar ranges using robust clipping at the 2nd and the 98th percentiles. Because our model ignores observational uncertainties, measure (b) may serve as a proxy for excess variance, another standard AGN variability measure; however, the actual values are overestimated and thus are not comparable with the literature, since we have not incorporated error bars. {\it Comments}: Duration changes smoothly across the AGN latent space, like in Figure~\ref{fig:latent-space-visualization-general}. On average, light curves with a larger magnitude variation are separated from those with smaller magnitude variations. The average colour shows a weak separation between redder and bluer AGNs, \reviewMoller{and similarly the peak colour shows separation from the reddest AGNs at peak from others}. \reviewMoller{The colour at the faintest ends of the light curve also shows weaker patterns. As a result, their difference shows that AGNs that are bluer when brighter are separated from others, and is in accordance with the commonly seen ``bluer-when-brighter'' trend}. In Appendix~\ref{appn:latent-space-std-features}, we show the $1\sigma$ standard deviation of features in each bin of the AGN latent space and generally find non-negligible scatter in bins corresponding to low-duration, high-variation AGNs, or those corresponding to considerably red and blue colors. Nonetheless, the trends in these plots are used in Section~\ref{subsubsec:agn-analysis} to roughly understand the positions of AGN light curves from Figure~\ref{fig:latent-space-only-agn-class-colored}.}\label{fig:latent-space-only-agn-features}
\end{figure*}

We correct the AGN light curves at test time before running the inference using the ALeRCE light curve correction pipeline\footnote{\url{https://github.com/alercebroker/lc_correction}} \citep[Appendix A of][]{Forster2021}. The primary motivation for accounting for the reference flux through this correction is to reveal the actual apparent magnitudes for these non-transients. However, the reference magnitude remained relatively constant for many AGNs\footnote{This is true for AGNs whose reference image was obtained in a very bright or faint epoch compared to the other epochs. However, if the template is at the middle of the magnitude distribution, it will significantly change the resulting light curve.} and thus this correction should not have a major effect on the model's test-time results because the correction renders the normalised magnitudes (Section~\ref{subsubsec:training}) agnostic to it. Out of the 2847 AGNs in the test set, 2359 had their light curves corrected, and 488 had their light curves uncorrected due to at least one of their alerts having a reference source farther than 1$\arcsec$.4 or due to divergence in corrected magnitude calculation. Only the corrected AGNs are considered in the subsequent discussion for consistency.

For this analysis, we exclude AGNs with durations shorter than 3 months. For the class-specific discussion based on Figure~\ref{fig:latent-space-agn}, we additionally exclude objects for which a Milliquas classification was not found (the `Unknown' class; Section~\ref{subsubsec:data-collection}).

We now interpret the positions of AGNs in the latent space using the spatial trends in Figure~\ref{fig:latent-space-only-agn-features}. However, there are two caveats. First, AGNs in some regions of latent space show noticeable scatter in light curve features, and possible reasons for this are discussed in Appendix~\ref{appn:latent-space-std-features}. Second, Milliquas uses luminosity-based criteria for separating AGNs from QSOs \citep[][]{Flesch2015}, but since our model does not use redshifts, some overlap among these classes is expected \citep[see also Appendix C.1 of][]{Sanchez_2021}. For these reasons, we focus on a broader latent space structure rather than a sharp class distinction.

{\it (A)}: \reviewMoller{Figure~\ref{fig:agn-light-curves-grid} highlights a handful of AGNs lying closer to SNe in the latent space than to the bulk AGN subpopulation\footnote{We verified that these AGNs remain separated from other AGNs across different UMAP runs with different seeds.}. As these regions are scarcely populated, this precludes the use of AGN features to discuss their location.}

\reviewMoller{Remarkably, AGNs with durations longer than nearly all SNe in our sample are also placed here (e.g., AGNs 1, 2, 3). This suggests that their photometric evolution over the temporal interval shared with nearby SNe is sufficiently similar to produce similar encodings.
Nonetheless, such contamination is rare, and a detailed characterisation is left for future work.}

{\it (B)}: \reviewMoller{These AGNs are characterised by high $\sigma/\mu$ (variability) and long durations (Figure~\ref{fig:latent-space-only-agn-features}), and contain the majority of BL Lacs in our sample (Figure~\ref{fig:latent-space-only-agn-class-colored}), consistent with their characteristic rapid variability. Very short blazar flares -- such as the $\approx$3-day flare in AGN4 near $t \approx 310$-340 days -- are not recovered by the interpolation. Longer flares spanning $\sim1$ to a few weeks -- such as those in AGN5 (around $t \sim 200$ d), AGN6 (just after $t \sim 300$ d), and AGN7 (at around $t \sim 450$ d) -- are captured more reliably but still imperfectly. These reflect challenges in encoding the overdensity of observations during rapid events.}

{\it (C)}: \reviewMoller{This region is characterised by lower variability and shorter durations on average compared to region B (Figure~\ref{fig:latent-space-only-agn-features}), though with large scatter in duration (Figure~\ref{fig:latent-space-only-agn-features-dur-std}). Its distinction from region A likely reflects the fact that they have low variability or even large peak times, atypical of our SNe sample, which is mostly declining.} BL Lacs \reviewMoller{AGN8} and \reviewMoller{AGN9} are unusually optically quiescent and therefore lie in this low-variability region. 

{\it (D)}: Region D shares the long-duration characteristic of region B but has lower variability on average. \reviewMoller{AGN10 is a BL Lac whose $\sigma/\mu$ and duration is comparable to region-B AGNs. This discrepancy remains unclear, but it might occur because its large $\sigma/\mu$ is driven by its significant rising trend rather than the rapid flaring characteristic of region B.}

\reviewMoller{To conclude, we have elucidated through a few examples that the AGN latent space correlates with duration, variability, and colour properties, and discussed how a tiny fraction of AGNs might reside closer to SNe due to larger similarities in their encodings than to other AGNs. This does not suggest that such AGNs are necessarily SNe-like, as that would require checking sensitivity to the local neighbourhood size in the UMAP projection (see Figure~\ref{fig:latent-space-visualization-general}'s caption). Instead, it is likely a consequence of using alert light curves, where the $5\sigma$ detection threshold can influence the observed durations and gaps in AGN light curves. As a result, duration may not be a relevant parameter for astrophysical interpretation of AGNs, even though our model captures it.}

\subsection{Visualisation of attention maps}\label{subsec:interpret}

The mTAND model used in this paper is trained to learn to attend to irregularly spaced observations (key) and produce interpolations over a set of regularly spaced query points (query). The corresponding attention weights are denoted by $\kappa$ in Equation~\ref{eqn:attn-weights}. In this section, we use the attention weights learned by the embedding function of the mTAND encoder (we use a single embedding function, $H = 1$; Section~\ref{subsubsec:training}) to understand how its outputs attend to observations. These are the outputs of the attention blocks (before the linear combination that produces the mTAN embedding) shown in Figure~\ref{fig:mtan-schema}.

\begin{figure*}
    \centering
    \includegraphics[width=0.4\linewidth,keepaspectratio]{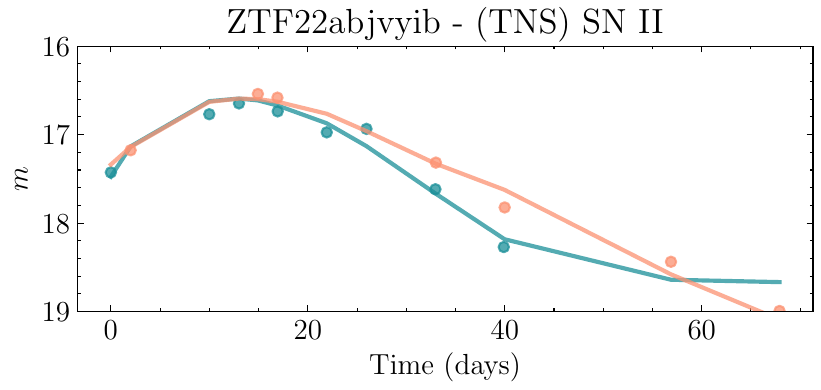}
    \includegraphics[width=0.4\linewidth,keepaspectratio]{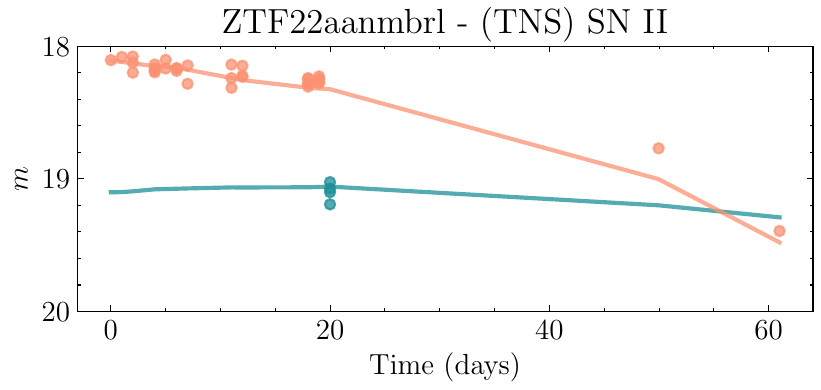}
    \includegraphics[width=0.4\linewidth,keepaspectratio]{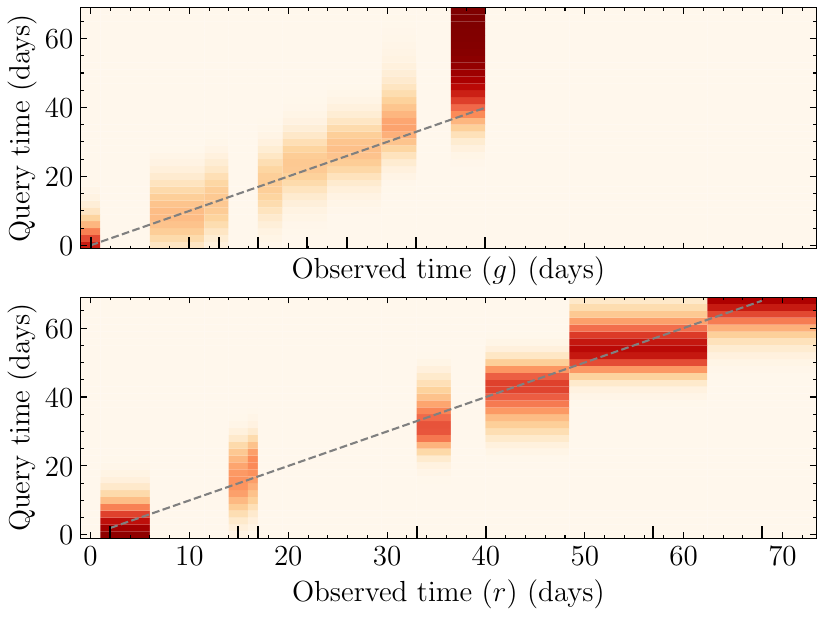}
    \includegraphics[width=0.4\linewidth,keepaspectratio]{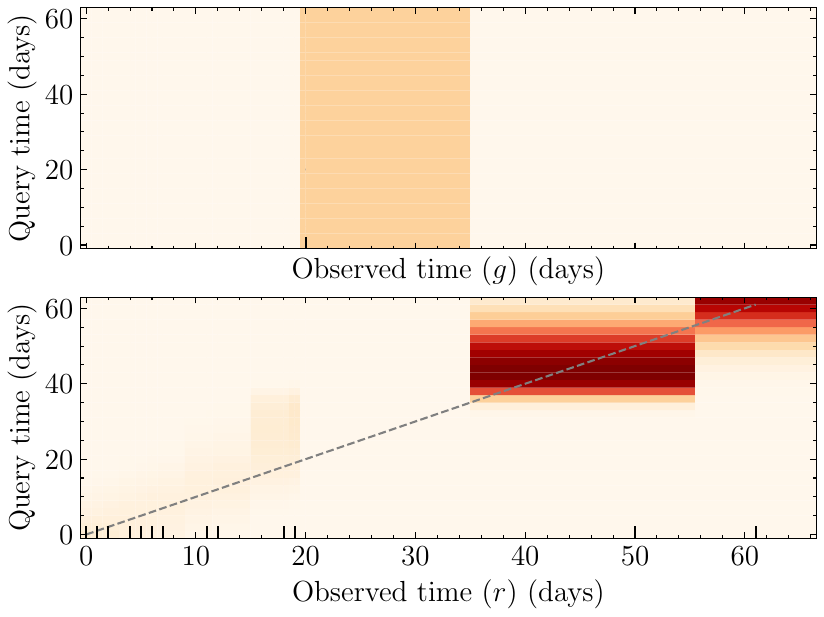}
    \includegraphics[width=0.4\linewidth,keepaspectratio]{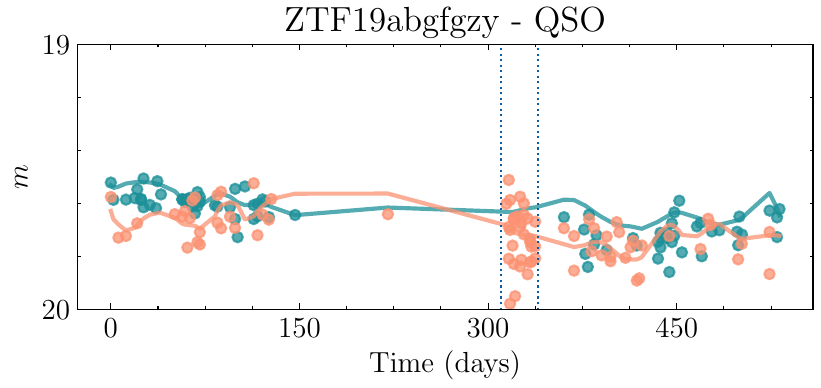}
    \includegraphics[width=0.4\linewidth,keepaspectratio]{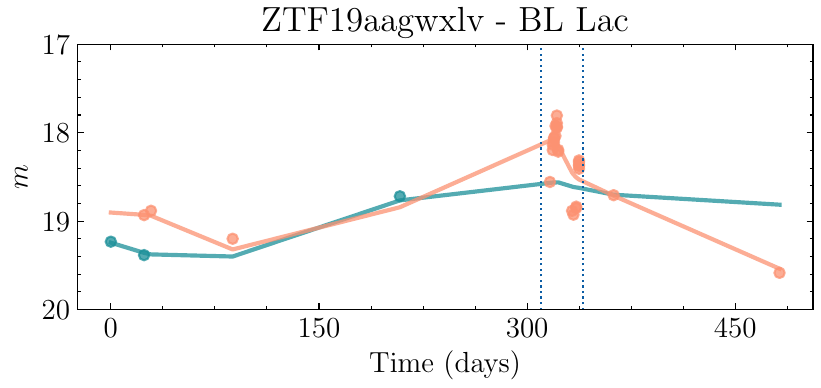}
    \includegraphics[width=0.4\linewidth,keepaspectratio]{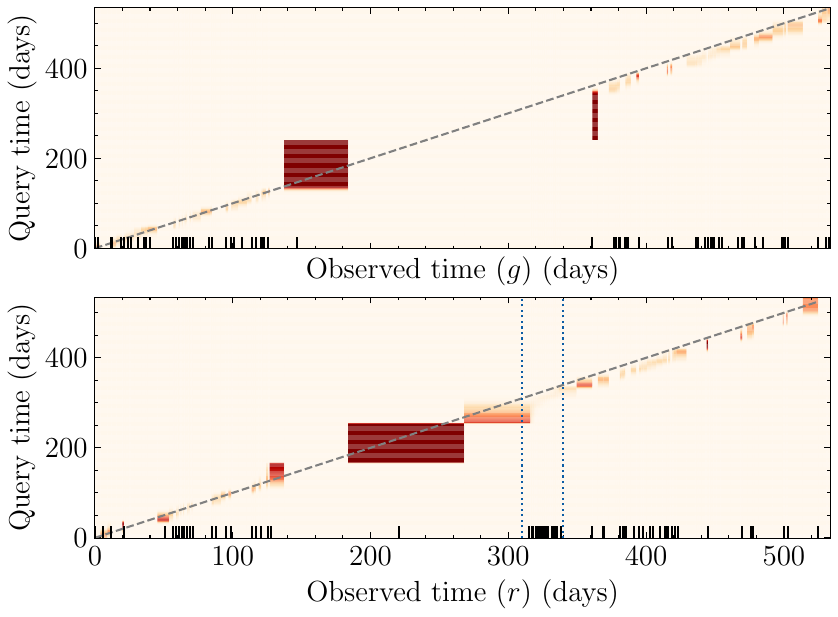}
    \includegraphics[width=0.4\linewidth,keepaspectratio]{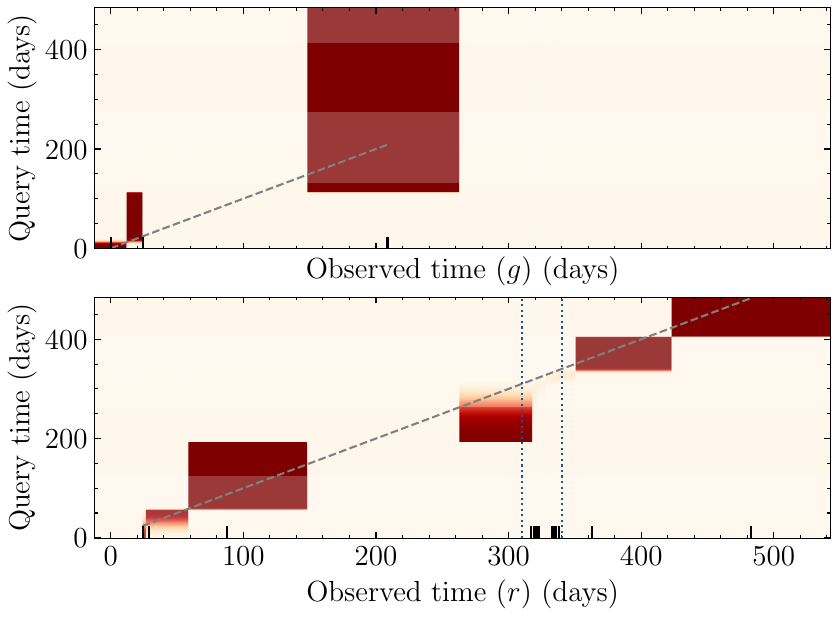}
    \caption{Example visualisation of attention weights for four light curves, where each panel shows the light curve followed by the attention maps for the $g$ and $r$ bands in the next two rows. Each attention map shows the (irregularly spaced) observed times on the x-axis and the (regularly spaced) query times on the y-axis. For reference, the vertical ticks on the x-axis denote the time at which an observation was present, and the dashed lines denote the $y = x$ line. The corresponding observed light curves, along with our model interpolations, are shown below each pair of attention maps. It is important to note that mTAND only attends to locations where an observation exists. The shading between `blocks' in the attention map is such that the boundaries of the blocks are defined at the midpoint of adjacent observation times. As a result, the horizontal width of the blocks reflects irregular spacing between observations, and large-width blocks should not necessarily be interpreted as a query attending to a large number of observations.}
    \label{fig:attn-maps}
\end{figure*}

Figure~\ref{fig:attn-maps} visualises the attention weight maps for each band separately for a few example light curves. Their visibly sparse and `diagonal'-like pattern suggests that queries attend to temporally nearby regions--this is exactly what is expected of mTAND, as longer-range dependencies are captured in its latent representation through the recurrent network following mTAND's encoder.

The attention weights for the two bands display distinct patterns, in that a given query point attends to different temporal regions in each band, depending on the location of observations for that band, which illustrates how each dimension/band of the light curve is uniquely represented on the set of query times. These intermediate representations of light curves are linearly combined using Equation~\ref{eqn:mtan-eqn} to obtain the mTAND encoder's latent representation, which thus explains how mTAN learns cross-band information.

Finally, this exercise can help understand model failure modes by analyzing any correspondence between attention weights and per-observation interpolation quality. Here, we briefly mention that fast-evolving observed features that tend to be missed by our interpolations receive lower attention. For example, this can be seen for ZTF19abgfgzy and ZTF19aagwxlv at $t \approx 310-340$ days in the $r$-band, where the attention weights are diminished (see the vertical dotted lines).

\subsection{Comparison with GP regression}\label{subsec:gp-reg-compare}

\begin{figure*}
    \centering
    \includegraphics[width=0.32\linewidth,keepaspectratio]{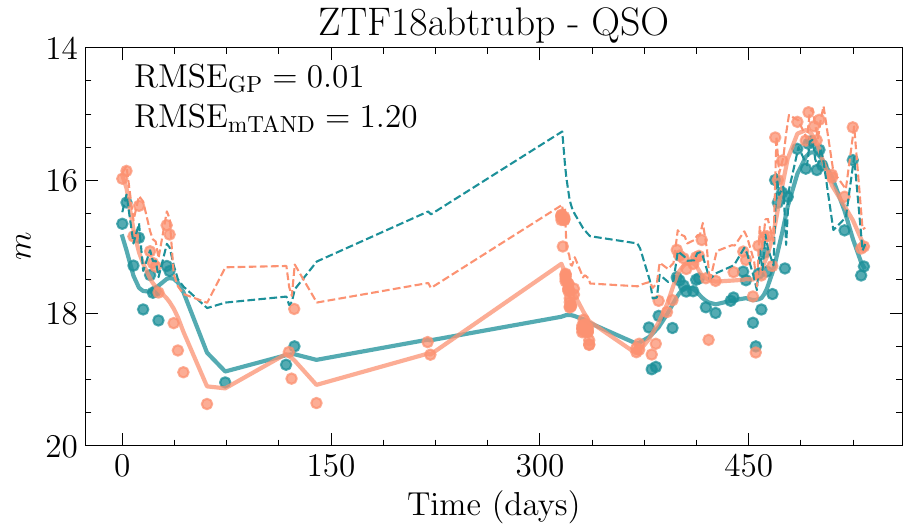}
    \includegraphics[width=0.32\linewidth,keepaspectratio]{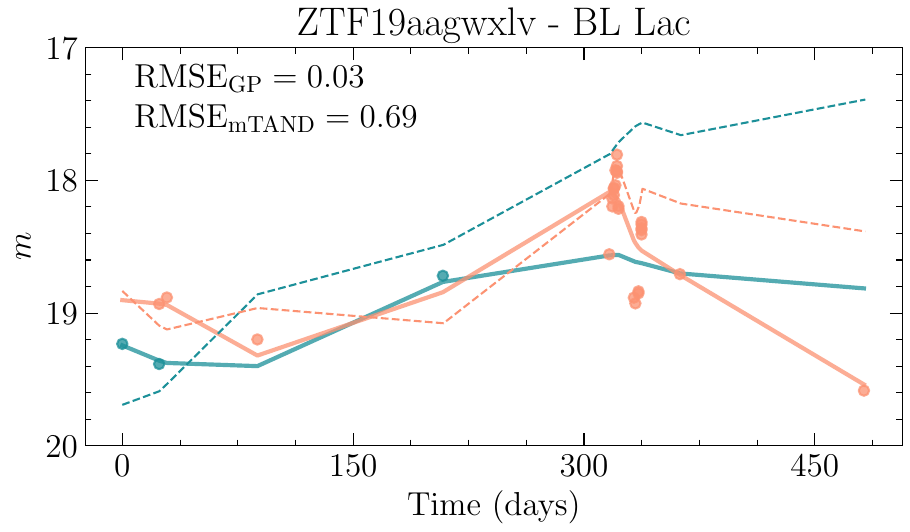}
    \includegraphics[width=0.32\linewidth,keepaspectratio]{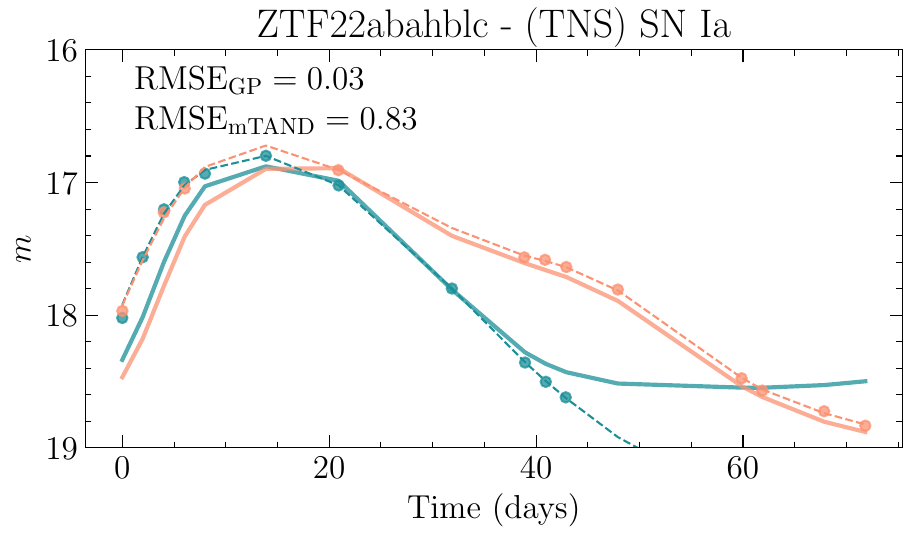}

    \caption{Examples of interpolations using mTAND and GP regression. The mTAND predictions are shown by solid lines and the mean GP predictions by dashed lines for $g$- (blue curves) and $r$-bands (red curves). The RMSEs for both approaches are denoted in each panel and are calculated using predictions from both bands.}
    \label{fig:mtan-vs-gp}
\end{figure*}

\begin{figure*}
\centering
\includegraphics[width=0.35\linewidth,keepaspectratio]{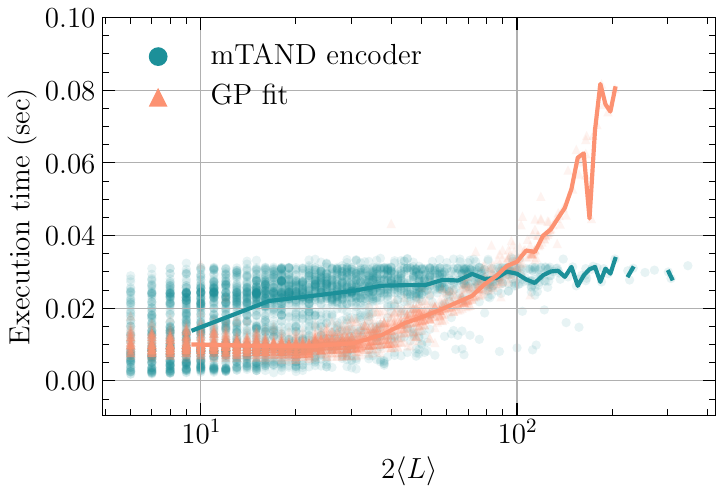}
\includegraphics[width=0.35\linewidth,keepaspectratio]{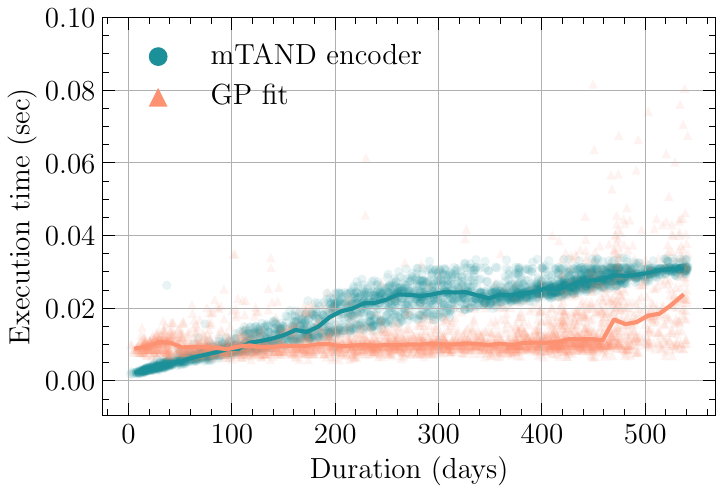}
\includegraphics[width=0.35\linewidth,keepaspectratio]{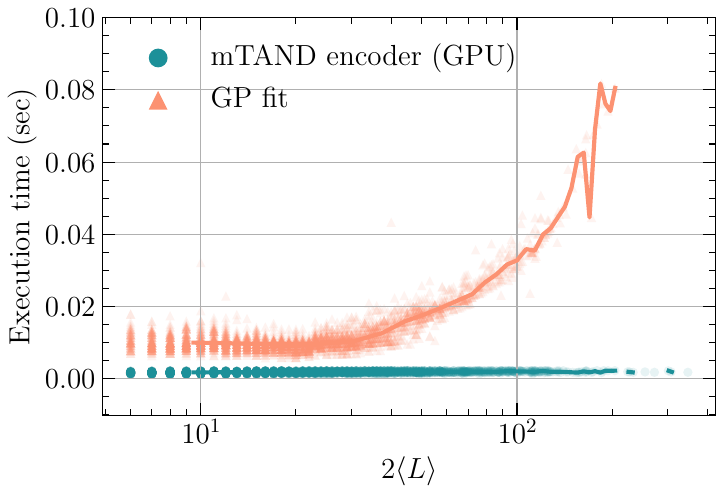}
\includegraphics[width=0.35\linewidth,keepaspectratio]{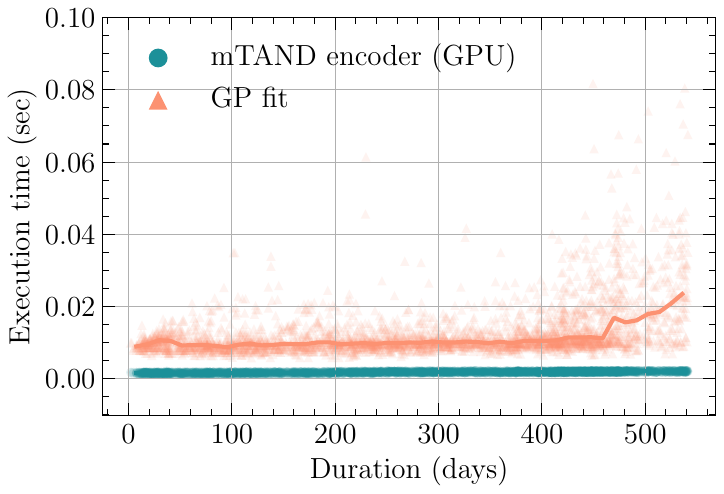}
\caption{Execution time vs. total no. of observations in the light curve (left) and duration (right), for the mTAND encoder (includes design of light-curve-dependent query times and encoder forward pass) and for the GP regression fit, for all light curves in the test dataset. The first and second rows show CPU and GPU times for mTAND, respectively. The solid lines denote the running median. Here, we use a batch size of one for the mTAND predictions. For clarity, the y-axis limits are truncated to remove some outlier observations; these remove some GP examples but do not affect the discussion of results.}\label{fig:mtan_vs_gp_per_light_curve}
\end{figure*}

We implement GP regression in both wavelength and time to handle cross-band correlations, following the approach of \citet{Boone_2019} and the code implementation of \citet{Allam_Jr_astronet_Multivariate_Time-Series_2022}. We use a Matérn 3/2 kernel, and evaluate GP predictions at the observed times\footnote{\reviewMoller{Although not shown here, evaluating the GP predictions on a finer uniform time grid spaced apart by two days (equivalent to mTAND's temporal resolution) produces spurious features at observational gaps for the two AGNs.}}. \reviewMoller{Although GP regression explicitly accounts for observational uncertainties, mTAND currently does not. We therefore use RMSE rather than the commonly used $\chi^2$ metric since it can be too strict for mTAND at the brightest ends that have small uncertainties. mTAND removes the restriction of GPs requiring a positive semidefinite covariance function, making it more flexible by design \citep{HetVAE_paper}.}

Figure~\ref{fig:mtan-vs-gp} compares the interpolation performance of mTAND and GP regression on three test-set light curves: two AGNs, one densely sampled and the other sparsely sampled, and an SN. GP regression produces better predictions than mTAND \reviewMoller{as shown by its smaller RMSE values}, particularly for outlier observations. This is a consequence of the fact that the GP fits are optimised for each light curve, whereas mTAND learns a single fit model for all light curves; this could also explain why mTAND does not model outlier observations as well. On the entire test set, the median RMSEs for GP and mTAND are 0.03 and 0.26, respectively, with median absolute deviations of 0.01 and 0.09.

We also discuss computational costs. As our goal is to characterise and classify light curves, we compare the execution time required for constructing (light curve-dependent) query times and encoding a light curve using mTAND versus performing a GP regression fit, both measured in identical CPU environments: Intel(R) Xeon(R) Gold 6140 CPU at 2.30GHz. For mTAND, we use a batch size of 32 to leverage its parallel processing capability, though we use a single process for loading data. mTAND requires 36 seconds (one minute with a batch size of one), and GP regression requires 41 seconds to execute its operations on the entire test set consisting of 3131 light curves; these values are the median over five independent runs. The similar runtimes of both GP and mTAND arise because, although exact GPs scale poorly with the number of observations ($O(n^3)$), most of our light curves have very few observations where GP costs remain reasonable. To investigate this further, we show in Figure~\ref{fig:mtan_vs_gp_per_light_curve} that GP indeed scales poorly with the observation count, whereas the execution time of mTAND remains unaffected by it.

The top-left panel of the figure also shows that mTAND is slower than GP on CPU for light curves with fewer than $\sim$80 observations, with a median time for such light curves of 0.023 versus 0.01 seconds and a median absolute deviation of 0.006 and 0.001 seconds, respectively. The top-right panel of Figure~\ref{fig:mtan_vs_gp_per_light_curve} shows that the execution time shows a mild dependence on light curve duration, which might be because mTAND learns time-representative encodings that can depend on the time span of the light curve. There are a considerable number of long-duration and sparsely sampled light curves (Figure~\ref{fig:data-stats}), and, taken together, these trends explain why mTAND is slower for light curves with few observations: it is driven by their also long duration. For all other combinations of observation count and duration (low and high), we find the mTAND to be at least slightly faster than GP. For example, for light curves with <80 observations and <120-day durations, the median runtimes of mTAND and GP are 0.006 and 0.009 seconds, with median absolute deviations of 0.002 and 0.001 seconds, respectively.

However, the bottom panels show that, on the GPU, mTAND is faster than GP across all light curves, including those with very few observations, and that its execution-time dependence on duration is diminished. On a per-light-curve basis, it is at least a few (5-10) times faster than on CPU; with a batch size of 32 across the entire test set, it takes just under a second, compared with 36 seconds on CPU. The GPU used is Tesla P100.

\subsection{\reviewMoller{Application to new variables and transients}}\label{subsec:OOD-test}

Here, we test the generalisability of our model by applying it to light curves with types not included in our training dataset. \reviewMoller{Specifically}, we apply our model to RR Lyrae stars, long-period variable stars, and TDEs, and briefly discuss their interpolation performance and position in the model's latent space. For a fair evaluation of the model's generalisation capabilities to new classes, the observed light curves are truncated so as not to exceed their duration compared to the light curves seen during training.

\begin{figure*}
    \centering
    \includegraphics[width=0.32\linewidth,keepaspectratio]{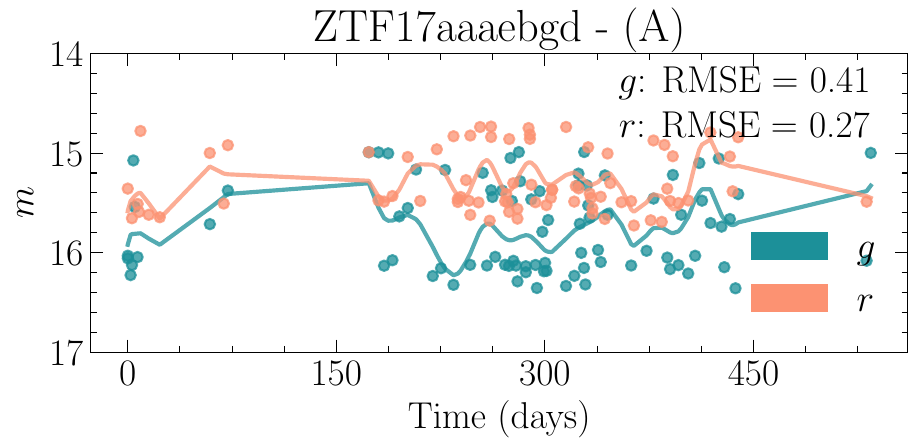}
    \includegraphics[width=0.32\linewidth,keepaspectratio]{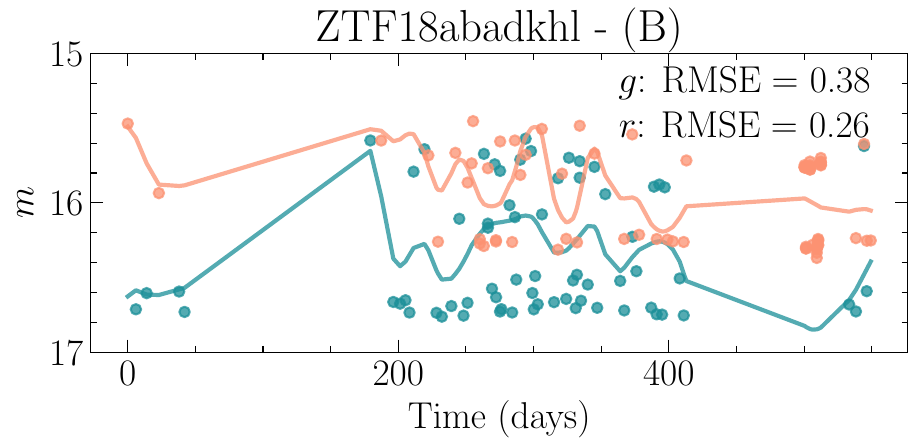}
    \includegraphics[width=0.32\linewidth,keepaspectratio]{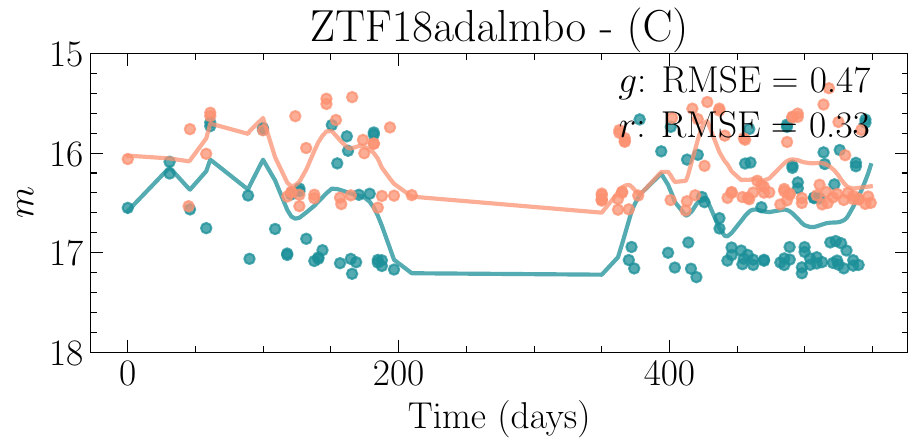}
    \includegraphics[width=0.32\linewidth,keepaspectratio]{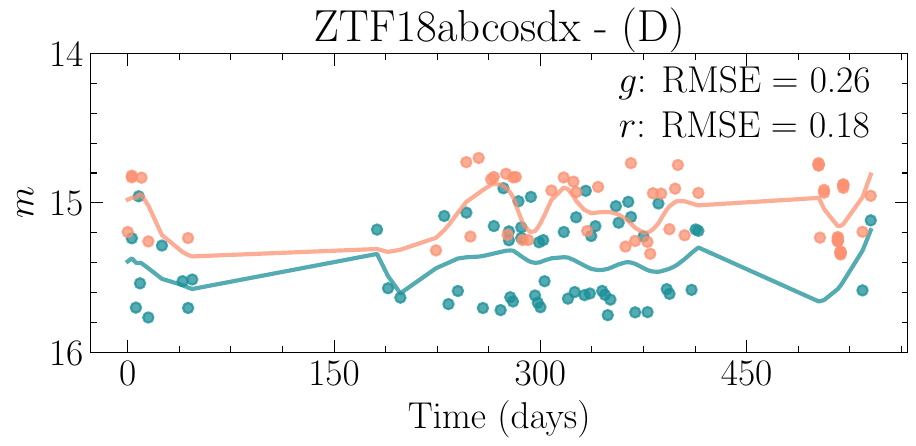}
    \includegraphics[width=0.32\linewidth,keepaspectratio]{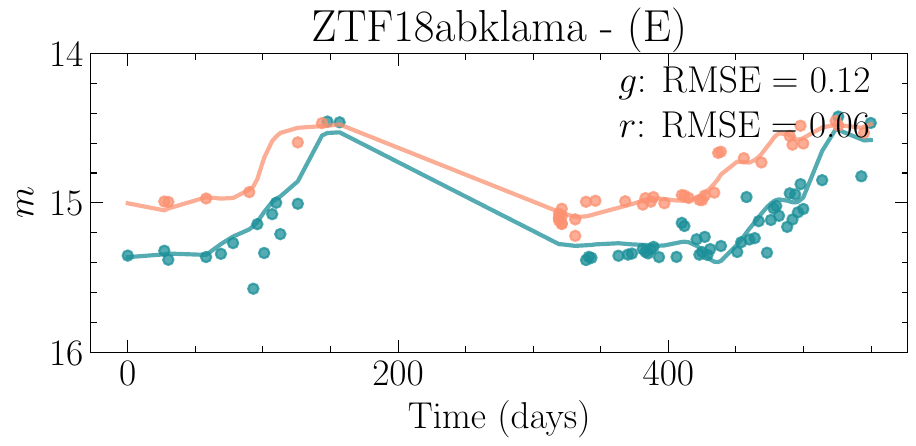}
    \includegraphics[width=0.32\linewidth,keepaspectratio]{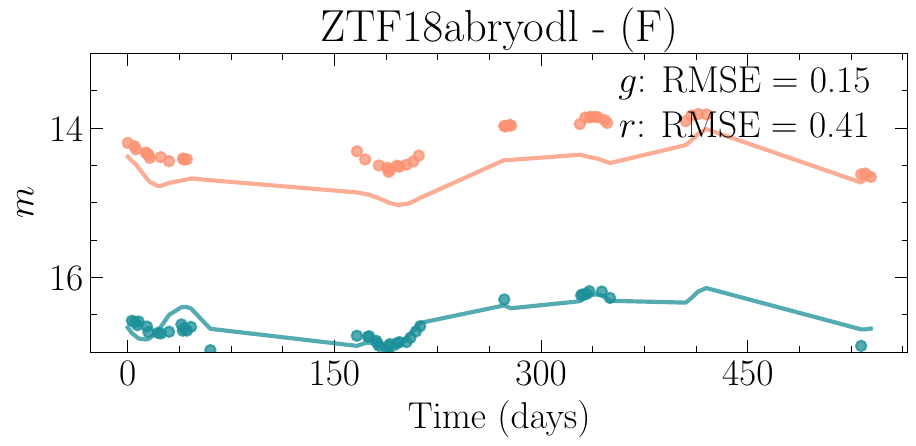}
    \includegraphics[width=0.32\linewidth,keepaspectratio]{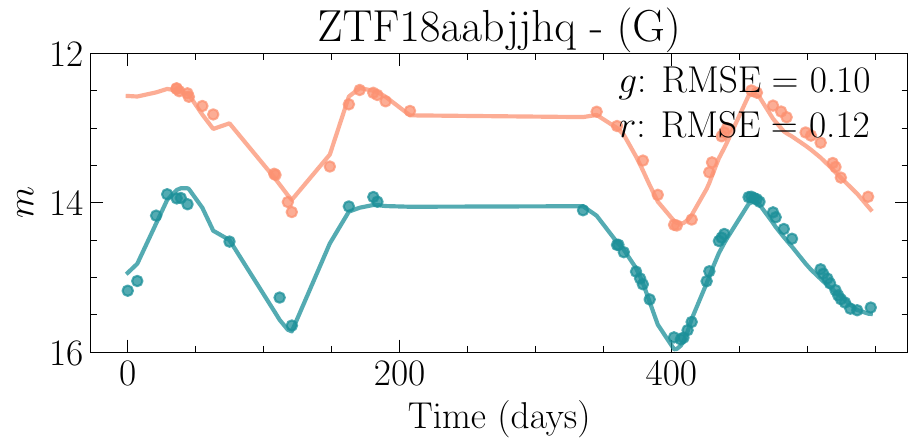}
    \includegraphics[width=0.32\linewidth,keepaspectratio]{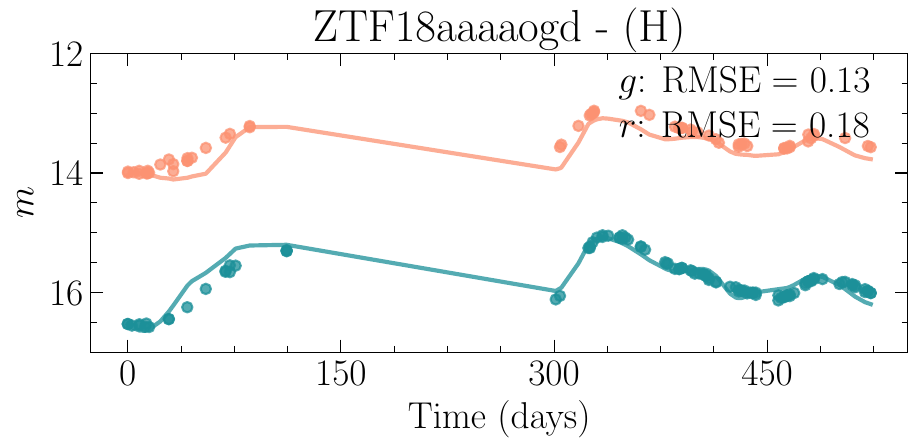}
    \includegraphics[width=0.32\linewidth,keepaspectratio]{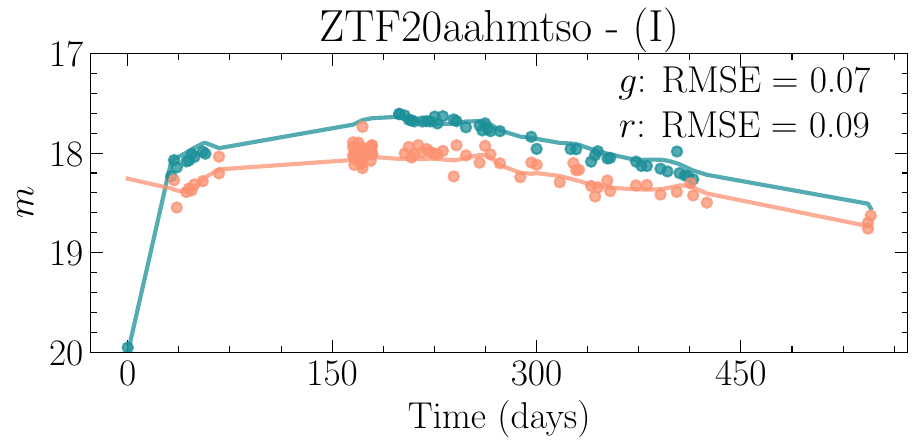}
    \includegraphics[width=0.32\linewidth,keepaspectratio]{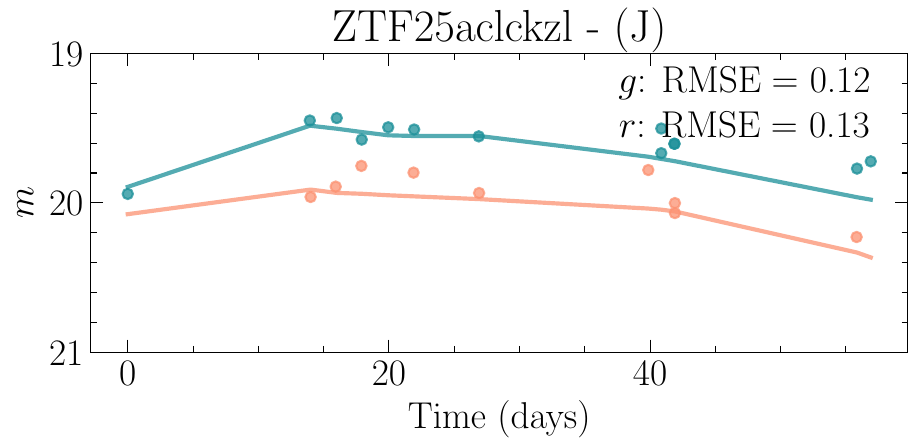}
    \includegraphics[width=0.32\linewidth,keepaspectratio]{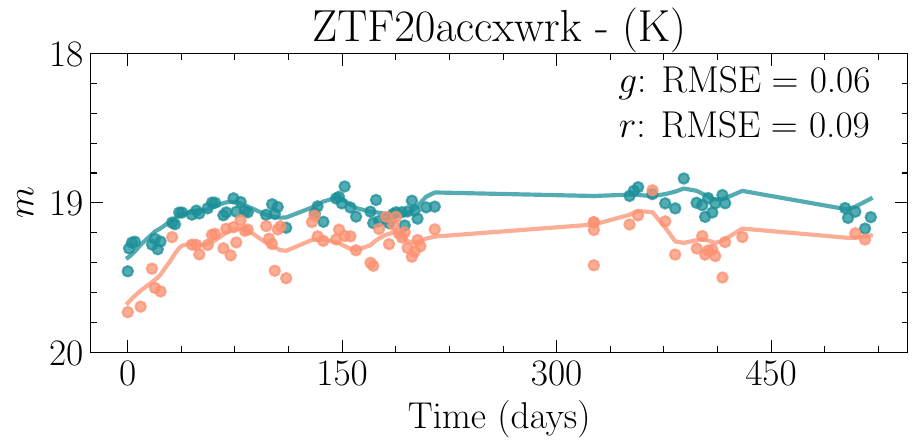}
    \includegraphics[width=0.32\linewidth,keepaspectratio]{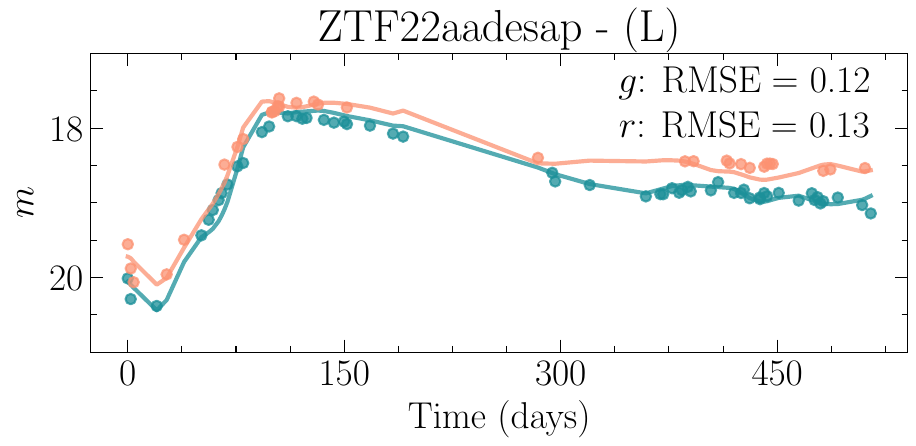}
    \includegraphics[width=0.32\linewidth,keepaspectratio]{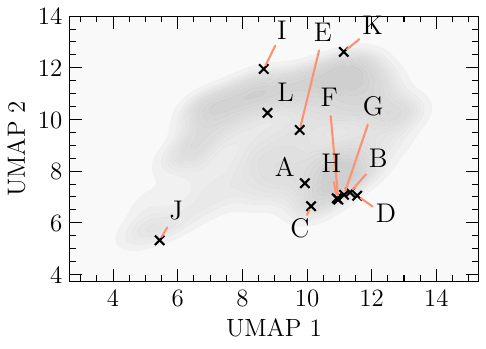}

    \caption{{\it Top}: Interpolation visualisations of the three \reviewMoller{types of} light curves with types not seen during training: RR Lyrae (\reviewMoller{A--E}), long-period variable (\reviewMoller{F--H}), and TDEs (\reviewMoller{I--L}). {\it Bottom}: The gray contour shows the latent space of SN and AGN (Section~\ref{subsec:latent-space}) and the crosses denote the locations of the three new light curves in the latent space.}
    \label{fig:OOD-interpolation-visualization}
\end{figure*}

Figure~\ref{fig:OOD-interpolation-visualization} shows the interpolations for the three \reviewMoller{types of} light curves and the corresponding locations our model positions them in the SNe + AGN latent space. The long-period variables and the TDEs show good decodings, with RMSEs similar to those seen in our previous analysis, despite these classes being absent from the training set. This generalisation is consistent with the fact that our model is trained to learn interpolations for arbitrary light curves in an unsupervised manner, rather than leveraging class-specific templates or prior information. The RR Lyrae show poor decoding performance for A--D, as the model could not capture its variability effectively--this is expected because its pulsation period (roughly 0.4-0.5 d) is much smaller than the temporal resolution of our model, which is two days.

Despite \reviewMoller{most of them having} similar durations, these sources occupy distinct regions in the latent space, which reconfirms that the model encodes variability patterns beyond duration alone. \reviewMoller{The RR Lyrae and the long-period variables lie away from SNe and towards AGNs, and that too (except for `E') preferably towards high-variability and redder AGNs}. For the long-period variable, this is understandable as they are known to be a source of contamination in variability-based AGN selection \citep[][]{MacLeod2011}. \reviewMoller{Unlike the long-period variables, RR Lyrae are not closely clustered in the latent space. Part of the reason might be their poor interpolation performance discussed above.} \reviewMoller{The TDEs are clearly separated from the variable stars. `I' and `K' lie at the edges of the AGN latent space, and `J' lies within the SN latent space, partly due to its small duration. The location of these TDEs is also consistent with their characteristic blue colour since they lie far away from redder light curves (see Figure~\ref{fig:latent-space-visualization-general}). We also checked that these lie far away from regions with a high variation in colour, which matches their typical behaviour of no colour evolution. Given the observational diversity of TDEs, it may be worthwhile to study how large samples of TDEs are arranged in the latent space.}

\section{Discussion}\label{sec:discussion}

\subsection{Summary and Outlook}

We have shown that our unsupervised mTAND approach is able to learn latent features that separate the general class of SNe and AGN populations from partial light curves obtained from ZTF, despite the significant class imbalance towards AGNs (90 vs. 10\%). In particular, the presence of correlations between positions of light curves in the latent space and derived features, such as peak magnitude epoch, light curve duration, magnitude variability, and overall and peak color, demonstrates that the latent space has learned multivariate photometric `properties' of our data. At the same time, the model is insensitive to unimportant features: peak and mean magnitude, and total number of observations. Duration, while informative for transient studies, can understandably be just an observational effect for objects such as AGNs, and since our latent space captures duration information, non-transient characterisation should be interpreted with this caveat in mind.

Within the SN subpopulation, duration was found to be one of the important features in organising the latent space. Different subtypes of SNe overlap in the latent space due to their similar photometric signatures. However, long-duration SNe II, early SN Ia candidates, and SLSNe were away from the generic SN subpopulation. The AGN latent space was less instructive than that of SNe due to more scatter in the relation between latent space location and light curve features--in fact, the finding by \citet{Sanchez_2021} that more than 30 features are required to separate among some AGN subclasses hints at this challenge. Moreover, as their spectroscopic classifications were based on luminosity, which the model did not have access to, the AGN analysis was focused on more general trends. A large fraction of Blazars were in regions of latent space corresponding to high variability. A few AGNs that were closer to SNe than the general AGN subpopulation were hypothesized to have SNe-like duration and variability or longer durations but similar variability across their temporal overlap; but this may also partly arise from the $5\sigma$ detection threshold used to trigger an alert, which can cause low-amplitude AGN variability to appear as transients.

We visualised the attention maps of the encoder network of mTAND and demonstrated that its multi-time attention mechanism captures the local structure of light curves, which is supplemented by the capture of longer-range temporal structure through a simple recurrent module following the encoder. By virtue of learning unique attention patterns for each band, we provide a high-level explanation of how mTAND learns cross-band information. We also found that our model's inability to interpolate over fast-evolving temporal features arises because the encoder places low attention on them. Unlike \citet{Pimentel_2023}, whose attention-based model focused on near-peak SNe observations (their Figure 12), our model does not seem to show such a preference. This difference may reflect our different data characteristics--even though both studies used ZTF light curves--or biases introduced by their pretraining on synthetic light curves where the peaks should be better defined than sparsely-sampled real observations.

\reviewMoller{GP regression provided more accurate interpolations than mTAND, including outlier/fast-evolving features, possibly because GP fits were optimised for each light curve. Thus, GP can be a good choice for analysing specific types of objects for follow-up studies where prompt characterisation is not a priority, because of its sound statistical grounding and its allowance of physically motivated kernel choices. However, we observed a few challenges with GP regression. First, they can produce spurious features at observational gaps that could bias downstream classification. Second, while kernel adjustments can improve GP interpolation accuracy, these modifications must result in a simple kernel to keep computational costs low, which can make GP suboptimal for handling a wide variety of light curves. mTAND learns a kernel from data, rather than being fixed, making it more flexible. Third, GP's computational scaling became prohibitive for $n \gtrsim$80 observations (Figure~\ref{fig:mtan_vs_gp_per_light_curve}), while mTAND's encoder runtime remained essentially insensitive to $n$. Therefore, mTAND can be particularly useful for large-scale applications, such as data from Rubin, having an overwhelmingly large dataset and six different bands. mTAND's runtime on CPU increased mildly with duration, likely due to its time-aware encoding, and hence it became slower than GP for long duration but sparse light curves; however, on GPU, mTAND showed no dependence of execution time on duration and remained faster than GP for all light curves.}

To demonstrate \reviewMoller{the generalisability of mTAN to light curves beyond those seen during training}, we applied the model to light curves of RR Lyrae, long-period variable stars, and TDEs. RR Lyrae's periodic variability occurred at a much faster rate than the temporal resolution of our mTAND of two days (which is a hyperparameter), and thus, mTAND could not model its variability well. \citet{Kelly2014} similarly found the CARMA model to be suboptimal for RR Lyrae, although the reason was the non-stochastic nature of these light curves, which resulted in model misspecification; mTAND, by contrast, is less sensitive to such assumptions. mTAND could, however, model the long-period variable and the TDE. The TDE was located in a sub-region of AGN latent space away from the redder light curves and those with large variability in colour, which matched the TDE's characteristic blue colour. Thus, the model could generalise to other transient classes by virtue of not using class-specific information during training. This aspect can be particularly valuable for rare transients from Rubin LSST. We note, however, that this generalisation assumes that test-set light curves have durations comparable to those in the training set, as we observed degraded interpolation for sections of light curve at times well beyond those seen during training ($\lesssim$1.5 years).

We note that we have developed and used a modified version of mTAND where the set of query (or reference) times is not fixed for all light curves but instead caters to each light curve independently. This approach was proven essential in our internal experiments to improve overall interpolation quality since, unlike the fixed query approach, it can capture finer temporal structures for light curves evolving over long timescales but also those on much shorter timescales.

mTAND's encoder required 36 seconds on a CPU but just under a second on a GPU to process $\sim$3k light curves. The \reviewMoller{model size} of the mTAND is only a few hundred kilobytes for both the encoder and decoder. Hence, the primary advantage of the mTAND is being fast and lightweight, while being useful for modelling a wide variety of transients. \reviewMoller{Direct comparison with similar methods is challenging due to differences in hardware (e.g., CPU cores used), runtime environments, and batch sizes. However, our mTAND encoder on CPU is $\sim$80 times faster than the `t2' time-series transformer model (at 1.5 seconds-per-light curve), even when evaluated with a batch size of one, and is roughly three times smaller \citep{Allam_2024}. A subsequent study \citep{Allam2023} reported significant improvements in that model, which reduced its size by $\sim$18 times and inference time by $\sim$8 using techniques such as deep compression and quantization; however, our current (unoptimised) implementation remains faster}. Finally, another advantage of mTAND is its temporally distributed latent space, unlike most contrastive learning approaches. Although we have not tested this idea, for example, this could open applications for phase-dependent analysis of light curves.

\subsection{Limitations and future directions}

The primary limitation is our neglect of observational uncertainties, which prevents the model from down-weighting noisier observations. At least two types of uncertainties can be considered. First are the uncertainties on the observed magnitudes, whose inclusion should be straightforward in our VAE architecture \citep[see e.g.,][]{ParSNIP_2021}. Second, uncertainties arising from temporal sparsity can, in principle, be captured by drawing multiple samples from our VAE latent space, but these uncertainties may not be adequately propagated through mTAN. An improvement to mTAN to allow representing a heteroscedastic distribution of outputs conditioned on the latent state, as suggested by \citet{HetVAE_paper}, could be worth exploring.

On a similar note, a more robust encoding can be obtained from our mTAND through multiple samplings from the posterior distribution over the latent states than a single sample used here. In addition, due to our assumption of a Gaussian distribution for the posterior, possible multimodalities in the true posterior distribution will not be captured, but we note that this is an inherent challenge for such neural networks as a tradeoff for the computational efficiency they provide. Finally, our data preprocessing made our mTAND less sensitive to observational effects that directly affect observed brightness, but a more general solution to this could be to incorporate such symmetries or other relevant host galaxy features in the model's latent space (section 3.3 of \citet{ParSNIP_2021} provides more details). This additional information can improve discrimination among transient subtypes more than our photometry-only approach, and even inform downstream parameter inference for huge samples of transients.

The second set of limitations arises from some characteristics of our model. As noted above, our interpolations struggle to capture temporal features evolving over timescales similar to or faster than the chosen temporal resolution of our mTAND. We find that with our two-day resolution, $\lesssim$1-week features were challenging to capture, and potentially a larger encoder latent dimension may be relevant to capture fast variations, for example, increasing the latent dimension from 2 to 32 only marginally increased the model memory footprint. Another idea for exploration is to exploit mTAN's flexibility by interfacing its encoder with alternative deep neural networks, such as convolutional or fully-connected networks, instead of the recurrent architecture used here. We acknowledge that rapid variability or short-lived events can be a defining characteristic of certain classes of transients, such as Blazars and early phases of some SNe, such as SN IIb and Ib/c. However, in this paper, we have shown that, at the level of data quality of ZTF, the mTAND is suitable for application to broad transient classification tasks where overall features are primary discriminants.

Another relevant task that can help in the efficient follow-up of transients is forecasting (extrapolation). In theory, this should be possible since the mTAND's decoder can be conditioned on any given query time, and because its time embedding contains a non-periodic term, which can capture evolution with time. However, the fully data-driven nature of our approach means that the extrapolations may be scientifically meaningful only when trained on focused datasets and applied to similar transient classes. Given the fast inference of the mTAND model, it may be worthwhile for future work to test its suitability for real-time follow-up using early-phase evolution \citep[see also, e.g.,][]{Sravan_2020,Muthukrishna2022}. For these purposes, the current mTAN architecture can, in principle, be modified to apply temporal masking to its decoder to base predictions only on past observations \citep[see][]{Vaswani2017}.

\reviewMoller{Finally, we leave for future work training for more epochs with data from other surveys, such as Rubin, particularly since our best model was obtained too close to the chosen end of our model training. We also leave for future work the application of mTAND to specific objects (e.g., AGNs). Although detailed physical characterisation is beyond the scope of this paper, future studies could study the organisation of latent space using astrophysical properties rather than potentially ambiguous classes (e.g., type IIs).}

\section{Conclusions}\label{sec:conclusion}

We have demonstrated a lightweight data-driven unsupervised representation learning method using a multi-time attention model (mTAND) that provides a scalable and flexible way of characterising irregular, partial light curves. We used alert data from the Fink broker, which is one of the Rubin/ZTF brokers \citep{fink_broker}. For transparency of evaluation, we only applied minimal selection cuts for photometric data quality with no heavy model-specific preprocessing. The main methodological modification to the mTAND we made was to allow the set of query times for the attention mechanism to fix the temporal resolution rather than to fix the number of query times. This is necessary to handle datasets containing light curves evolving on very different time scales.

Our approach learns latent representations correlating with physically meaningful features (duration, peak time, variability, colour) without supervision, while being robust to observed magnitude and observation count. By virtue of this, the model was able to separate SNe and AGN light curves as a whole despite heavy class imbalance. Expected photometric overlaps among subtypes of SNe and AGN and anomalies in the latent space were elucidated.

The mTAND produced good decodings and latent representations of light curves to long-period variables and TDEs, classes that were unseen during training. Application to RR Lyrae, however, did not produce reasonable decodings due to its periodic variability occurring much faster than our model’s temporal resolution. Overall, the model shows potential for application when training samples may not be representative of the whole population within a given survey.

The limitations in our current application of mTAND are the ignorance of observational uncertainties, the inability to adequately capture evolution faster than our model's temporal resolution, and lower interpolation accuracy than GP per-object fits. We have discussed how some of these challenges can be addressed within the framework of our model.

Fast, data-driven characterisation of transients is important for Rubin since it will detect a few million transients per night with large sky coverage and depth. Our model's scalability arises from a few aspects. It required just under a second on a GPU (around half a minute on CPU) to process $\sim$3k light curves, achieved execution times that remained invariant to the number of observations in the light curve (and also to duration when used on a GPU), and the model occupies only of the order of few hundred kilobytes of memory at runtime. The main benefits compared to GP regression are its ability to learn complex kernel functions in a data-driven manner, which mitigates the risk of model misspecification, and insensitivity of its runtime to observation count.

The results here suggest that mTAND can serve as an effective classifier to rapidly identify strong candidates for follow-up based on general photometric features, with precise subtype discrimination (supervised classification) and uncertainty quantification as natural next steps. As time-domain astronomy enters a regime dominated by sheer data volume, methods such as mTAND can be beneficial for new discoveries and characterisation of both common and rare transient populations whose physical understanding has thus far been elusive due to statistical errors arising from limited classified sample sizes.

\section*{Acknowledgements}

This work was performed on the OzSTAR national facility at Swinburne University of Technology. The OzSTAR program receives funding in part from the Astronomy National Collaborative Research Infrastructure Strategy (NCRIS) allocation provided by the Australian Government, and from the Victorian Higher Education State Investment Fund (VHESIF) provided by the Victorian Government.
This work was developed within the Fink community and made use of the Fink community broker resources. Fink is supported by LSST-France and CNRS/IN2P3. This research has made use of the SIMBAD database, operated at CDS, Strasbourg, France. AM research was supported by the Australian Research Council Discovery Early Research Award (DE230100055).
Y.G. acknowledges the STFC Centre for Doctoral Training in Data Intensive Science and support from the UCL Research Excellence Scholarship (UCL-RES). Y.G. also acknowledges Ofer Lahav for discussions and for providing additional comments.

\section*{Data Availability}

The code for designing and training the mTAN model, and other associated scripts, is available at \url{https://github.com/Yash-10/astro-mtan}. The data was obtained using Fink's Data Transfer service for ZTF: \url{https://ztf.fink-portal.org/download}, and is publicly available.



\bibliographystyle{mnras}
\bibliography{biblio} 




\appendix

\section{Latent space correlation patterns with extraneous light curve features}\label{appn:latent-space-extraneous}

\begin{figure*}
    \centering
    \includegraphics[width=0.32\linewidth,keepaspectratio]{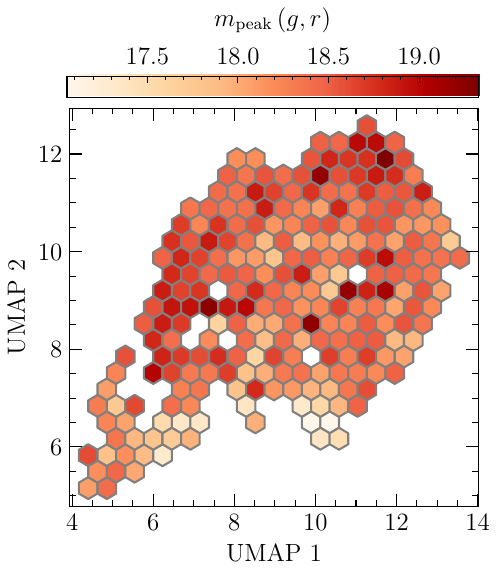}
    \includegraphics[width=0.32\linewidth,keepaspectratio]{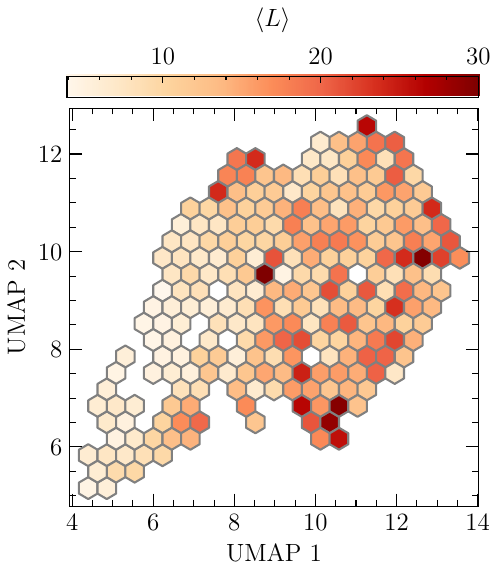}
    \includegraphics[width=0.32\linewidth,keepaspectratio]{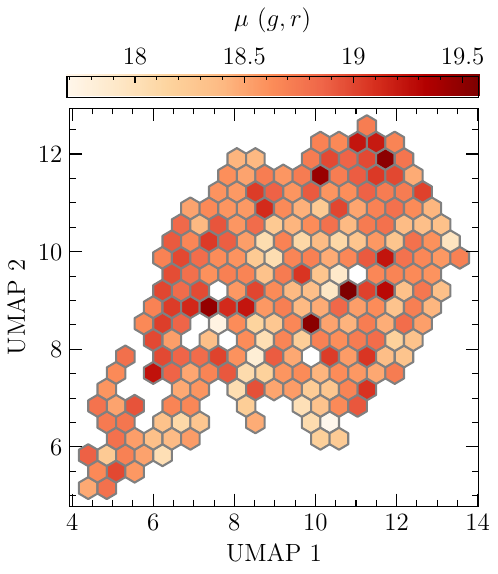}
    \caption{Same as Figure~\ref{fig:latent-space-visualization-general} but colored with different set of features. See Figure~\ref{fig:latent-space-visualization-general} caption for details. The features used to colour the latent space are: peak magnitude irrespective of the band ($m_{\mathrm{peak}}\,(g, r)$; left; this is used for normalising magnitudes before training), average no. of observations in each band ($\langle L \rangle$; middle; also used in Figure~\ref{fig:data-stats}, and the average of mean magnitude across both bands ($\mu\,\,(g, r)$; bottom)).}
    \label{fig:latent-space-visualization-general-extraneous}
\end{figure*}

As mentioned in Section~\ref{subsec:latent-space}, the model is not expected to encode the peak magnitude (calculated across all bands) because it is subtracted from light curve magnitudes before training (Section~\ref{subsubsec:training}), and the total number of observations across all bands, because the training loss function was normalised using this term (Section~\ref{subsubsec:encoder-decoder-framework}). In addition, we expect the model's latent space not to learn the mean magnitude. Like peak magnitude, being robust to mean magnitude is important, as this can be a misleading feature for classifying transients. Figure~\ref{fig:latent-space-visualization-general-extraneous} shows the visualisation of the latent space, similar to Figure~\ref{fig:latent-space-visualization-general}, but colored with these extraneous features.

\reviewMoller{The peak and mean magnitudes do not show obvious correlations apart from some preference seen for very bright peak magnitudes. The average number of observations indicates that light curves towards the left in the projection have fewer observations, which can be viewed as a surrogate for duration.}

\section{\reviewMoller{SN latent space correlations with remaining properties}}\label{appn:latent-remaining-features}

\begin{figure*}
    \begin{subfigure}{0.32\linewidth}
        \includegraphics[width=\linewidth,keepaspectratio]{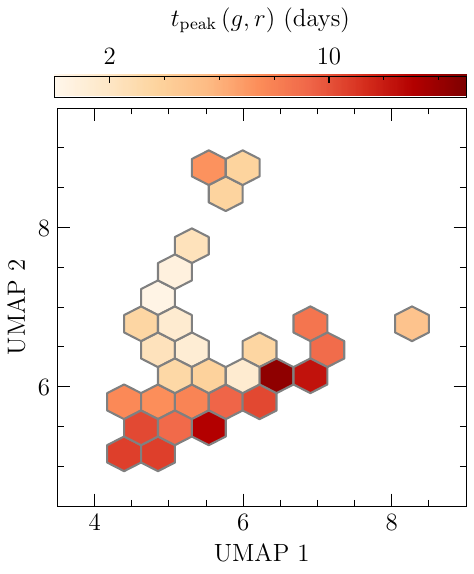}
        \caption{}\label{fig:latent-space-only-sn-features-tpeak}
    \end{subfigure}
    ~
    \begin{subfigure}{0.32\linewidth}
        \includegraphics[width=\linewidth,keepaspectratio]{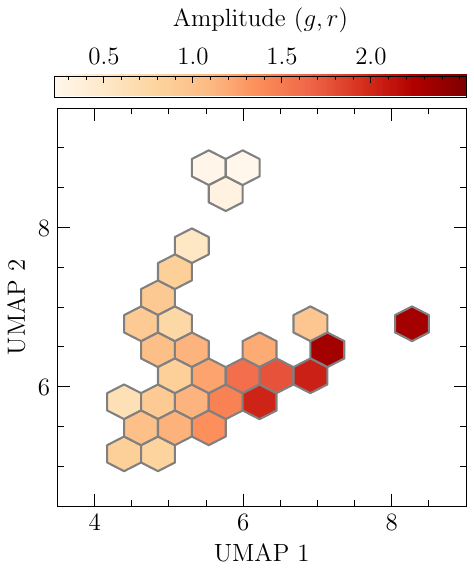}
        \caption{}\label{fig:latent-space-only-sn-features-amplitude}
    \end{subfigure}
    \caption{Same as Figure~\ref{fig:latent-space-only-sn-features} but colored with two other properties: peak time and amplitude.}\label{fig:latent-space-only-sn-features-remaining}
\end{figure*}

\reviewMoller{Figure~\ref{fig:latent-space-only-sn-features} showed the SN latent space colored by duration, decline rate, and colour at peak. Here in Figure~\ref{fig:latent-space-only-sn-features-remaining}, we show correlations with time of peak and amplitude.}

\reviewMoller{The peak time correlates spatially in the latent space, but note that our individual band peak times are sensitive to sampling and inaccurate in cases of light curves showing flatter peaks. The amplitude shows smooth correlation, and these patterns are qualitatively similar to the duration plot. The fact that the duration and amplitude plots show similar patterns is not unexpected, as the longer the SN evolution is observed, the more the variation can be captured. Although not shown, the standard deviation of colour ($\sigma_{\mathrm{g - r}}$) and the standard deviation of magnitudes show qualitatively similar patterns to the amplitude plot shown here, where the former may be capturing the tendency of SNe to redden during decline.}

\section{Standard deviation of features in latent space}\label{appn:latent-space-std-features}

\begin{figure*}
    \begin{subfigure}{0.32\linewidth}
        \includegraphics[width=\linewidth,keepaspectratio]{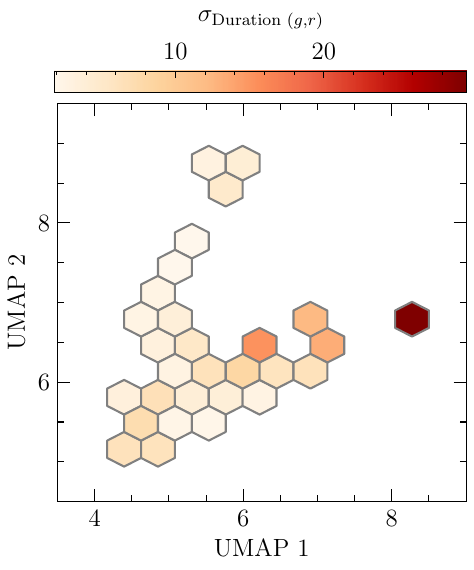}
        \caption{}\label{fig:latent-space-only-sn-features-dur-std}
    \end{subfigure}
    ~
    \begin{subfigure}{0.32\linewidth}
        \includegraphics[width=\linewidth,keepaspectratio]{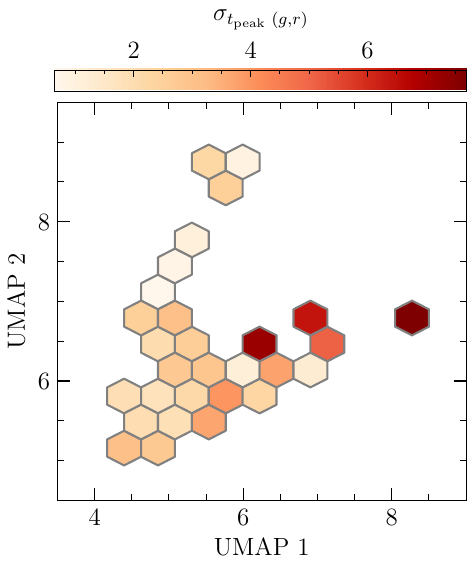}
        \caption{}\label{fig:latent-space-only-sn-features-tpeak-std}
    \end{subfigure}
    ~
    \begin{subfigure}{0.32\linewidth}
        \includegraphics[width=\linewidth,keepaspectratio]{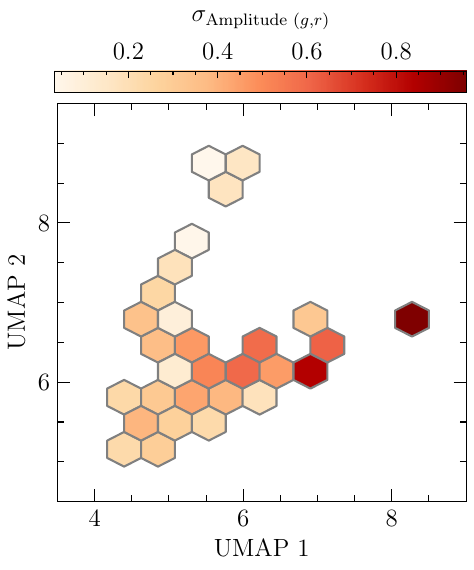}
        \caption{}\label{fig:latent-space-only-sn-features-amplitude-std}
    \end{subfigure}
    ~
    \begin{subfigure}{0.32\linewidth}
        \includegraphics[width=\linewidth,keepaspectratio]{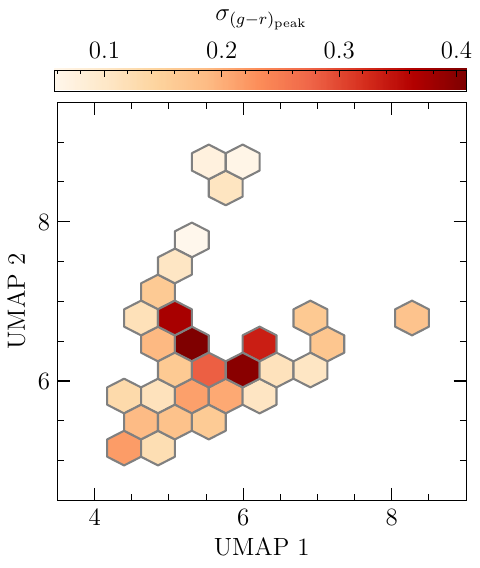}
        \caption{}\label{fig:latent-space-only-sn-features-colorpeak-std}
    \end{subfigure}
    \caption{}\label{fig:latent-space-only-sn-features-std}
\end{figure*}

Figure~\ref{fig:latent-space-only-sn-features-std} shows the latent space of SN using the same features as in Figure~\ref{fig:latent-space-only-sn-features} (except the magnitude decline rate), but instead of mean features in each bin, it denotes their $1\sigma$ standard deviation. This allows us to understand the model's `confidence' in clustering light curves based on these features. When compared with Figure~\ref{fig:latent-space-only-sn-features}, we observe that with respect to the mean values in the bins, the standard deviations are not large. In particular, while the standard deviation of colour is large at locations of the reddest SNe, it is small for mildly red and blue SNe, which means that they are separated with reasonable certainty. Thus, the model indeed clusters light curves with similar features with good confidence, and the mean trends in the main text are generally reliable.

\begin{figure*}
        \begin{subfigure}{0.32\linewidth}
        \includegraphics[width=\linewidth,keepaspectratio]{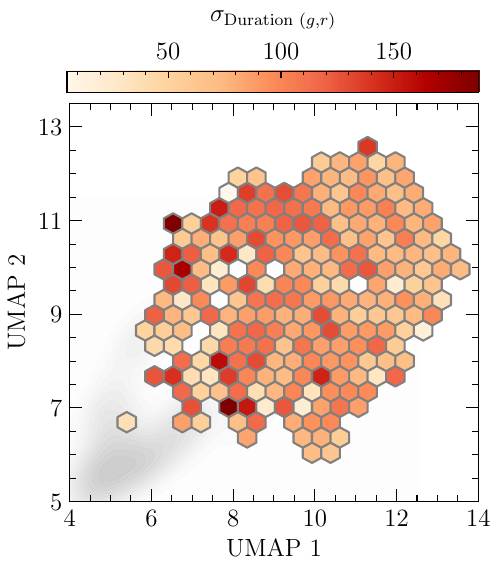}
        \caption{}\label{fig:latent-space-only-agn-features-dur-std}
    \end{subfigure}
    ~
    \begin{subfigure}{0.32\linewidth}
    	\includegraphics[width=\linewidth,keepaspectratio]{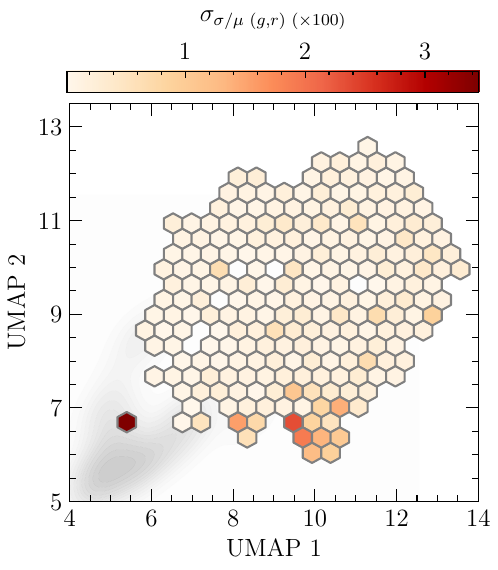}
    	\caption{}\label{fig:latent-space-only-agn-features-meanvariance-std}
    \end{subfigure}
    ~
    \begin{subfigure}{0.32\linewidth}
        \includegraphics[width=\linewidth,keepaspectratio]{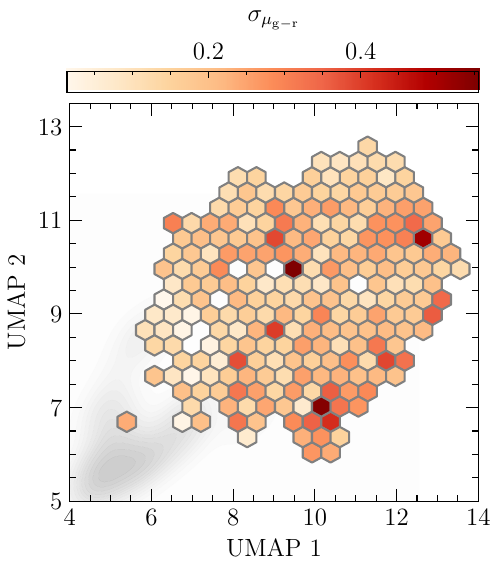}
        \caption{}\label{fig:latent-space-only-agn-features-meancolor-std}
    \end{subfigure}
    ~
    \begin{subfigure}{0.32\linewidth}
        \includegraphics[width=\linewidth,keepaspectratio]{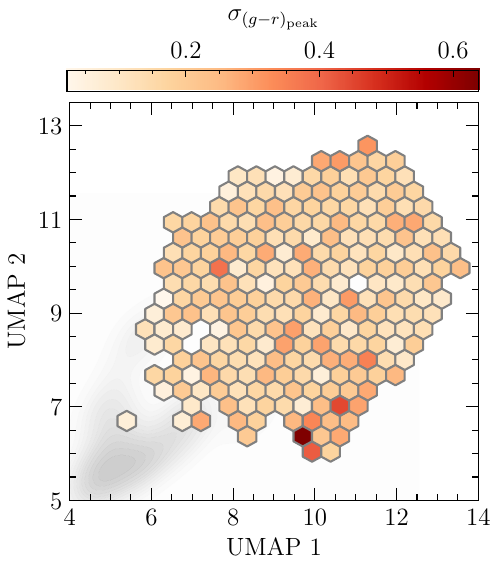}
        \caption{}\label{fig:latent-space-only-agn-features-peakcolor-std}
    \end{subfigure}
    ~
    \begin{subfigure}{0.32\linewidth}
        \includegraphics[width=\linewidth,keepaspectratio]{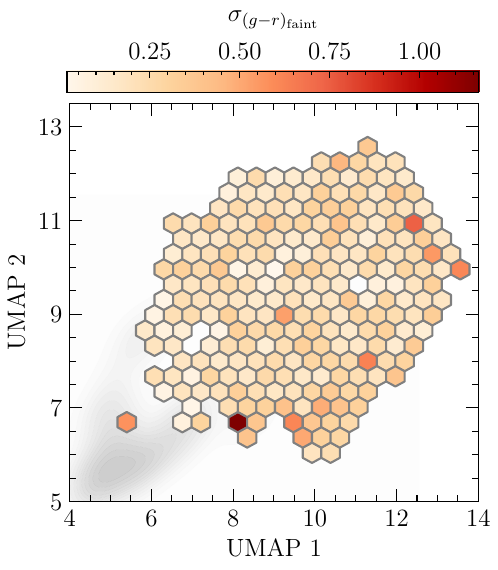}
        \caption{}\label{fig:latent-space-only-agn-features-faintestcolor-std}
    \end{subfigure}
    ~
    \begin{subfigure}{0.32\linewidth}
        \includegraphics[width=\linewidth,keepaspectratio]{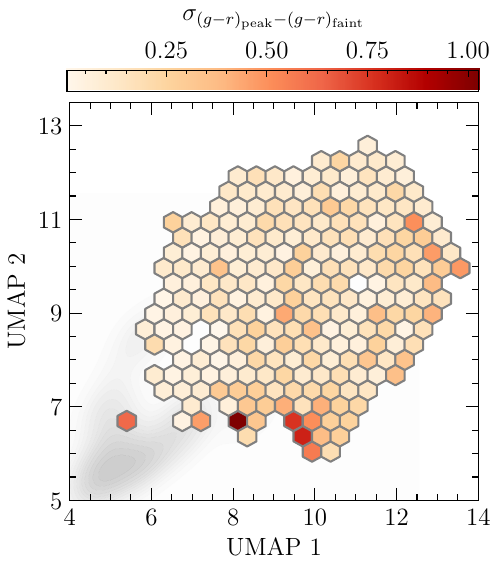}
        \caption{}\label{fig:latent-space-only-agn-features-peakminusfaintestcolor-std}
    \end{subfigure}
    \caption{}\label{fig:latent-space-only-agn-features-std}
\end{figure*}

Figure~\ref{fig:latent-space-only-agn-features-std} shows the standard deviation in each bin of the AGN latent space. Comparing it with Figure~\ref{fig:latent-space-only-agn-features}, we see that some bins defined by relatively low-duration AGN light curves (duration $\lesssim200$ days) have comparable or larger standard deviation than their duration, whereas the bins defined by the longest-duration AGN light curves have lower relative scatter. \reviewMoller{The standard deviation in the variability is generally of similar scale to the variability for both high and low variability regions (Figure~\ref{fig:latent-space-only-agn-features-meanvariance}).} \reviewMoller{Similarly, the standard deviation in the average colour is not small, particularly for the reddest and the bluest bins, and the standard deviation in the peak color, faintest color, and their difference are all considerable. There are following possible reasons for this non-negligible scatter: (a) it might occur because our model is trained to perform interpolations which try to `fit' as many points as possible, rather than smoother modeling, which can lead to this behaviour for (stochastic) AGNs unlike for (non-stochastic) SNe, (b) the hyperparameter setting in the UMAP projection: in particular, the tradeoff between the local and global structure can be controlled when performing the UMAP projection and a smaller neighbourhood size could also reduce the scatter}, (c) it might also reflect data characteristics: for example, high $\sigma/\mu$ can arise due to changing mean trend, high variation around the mean, or both, since $\mu$ here is the overall mean magnitude rather than a running mean. Thus, although these features may not individually be good descriptive features to understand the \reviewMoller{`local'} clustering of AGN light curves, the mean trends in Figure~\ref{fig:latent-space-only-agn-features} it is representative of what the model has learned generally on a \reviewMoller{`global' level}.

\bsp	
\label{lastpage}
\end{document}